\documentclass{aa}  
\usepackage{txfonts}
\begin{document}

\title{3D mapping of young stars in the solar neighbourhood \\ with \textit{Gaia} DR2}
\authorrunning{E. Zari et al.}
\titlerunning{Solar neighbourhood in 3D}
\author{E. Zari \inst{1}, H. Hashemi \inst{1},  A. G. A. Brown \inst{1},  K. Jardine \inst{2}, and  P.T. de Zeeuw \inst{1}
}
\institute{
{1} Leiden Observatory, Leiden University, Niels Bohrweg 2, 2333 CA Leiden, the Netherlands; \\
{2} Consultant, Radagast Solutions, Simon Vestdijkpad 24, 2321 WD Leiden, the Netherlands; \url{galaxymap.org}
}

\abstract{
We study the three dimensional arrangement of young stars in the solar neighbourhood using the second release of the \textit{Gaia} mission (\textit{Gaia} DR2) and we provide a new, original view of the spatial configuration of the star-forming regions within $500 \, \mathrm{pc}$ of the Sun. By smoothing the star distribution through a Gaussian filter, we construct three dimensional (3D) density maps for early-type stars (upper-main sequence, UMS) and pre-main sequence (PMS) sources.
The PMS and the UMS samples are selected through a combination of photometric and astrometric criteria. A side product of the analysis is a 3D, $G$-band extinction map, which we use to correct our colour-magnitude diagram for extinction and reddening.
Both density maps show three prominent structures, \textit{Scorpius-Centaurus, Orion,} and \textit{Vela}. The PMS map shows  a plethora of lower-mass star-forming regions, such as \textit{Taurus, Perseus, Cepheus, Cassiopeia,} and \textit{Lacerta}, which are less visible in the UMS map due to the lack of large numbers of bright, early-type stars.
We  report the finding of a candidate new open cluster towards $l, b \sim 218.5^{\circ}, -2^{\circ}$, which could be related to the Orion star-forming complex. We estimate ages for the PMS sample and we study the distribution of PMS stars as a function of their age. We find that younger stars cluster in dense, compact clumps, and are surrounded by older sources, whose distribution is instead more diffuse. 
The youngest groups that we find are mainly located in \textit{Scorpius-Centaurus, Orion, Vela,} and \textit{Taurus}. \textit{Cepheus, Cassiopeia,} and \textit{Lacerta} are instead more evolved and less numerous. Finally, we find that the 3D density maps show no evidence for the existence of the ring-like structure which is usually referred to as the Gould Belt. 
}


\titlerunning{solar neighbourhood DR2}

\maketitle

\section{Introduction}
Since the second half of the  nineteenth century, it was recognised by Herschel (1847) and Gould (1874) that the brightest stars are not distributed randomly in the sky, but seem to  form a belt (which afterwards became known as the Gould Belt) with an inclination of $\sim 20^{\circ}$ with respect to the plane of the Milky Way. Furthermore, O- and B-type stars clustered in loose groups that were named `associations' by Ambartsumian (1947). The Gould Belt was subsequently found to be associated with a significant amount of interstellar material \citep{Lindblad1967}, interpreted as an expanding ring of gas \citep{Olano1982, Elmegreen1982}. Giant molecular clouds were  also found to be related to the most prominent OB associations \citep{Sancisi1974, Kutner1977, deGeus1992, Dame1993}. This agrees well with the fact that OB associations are young, as supported by the ages derived from colour-magnitude diagrams. 

The origin of the Belt is debated, and various formation scenarios have been proposed. \cite{Comeron1992} and \cite{Comeron1998} proposed that the Gould Belt was formed after the oblique impact of a high-velocity cloud on the galactic disc. \cite{Poppel1997} suggested instead a cascade of supernova explosions. Alternatively, \cite{Olano2001} proposed that a supercloud of $2 \times 10^7 \mathrm{M_{\odot}}$ and  400 pc in size is the common precursor of the Sirius super cluster, the Gould Belt, and the Local Arm. The breaking and compression of the supercloud  would have produced the latter two, while the cluster, unaffected by friction would have moved on, away from the gas system.
Finally, \cite{Bekki2009} suggests that the Belt was formed after the collision between a gas cloud of $\sim 10^6 \mathrm{M_{\odot}}$ and a $\sim 10^7  \mathrm{M_{\odot}}$ dark matter clump, based on numerical simulations of the collision. 

Many studies have described the structure and the kinematics of the Gould Belt. Thanks to the data of the \textit{Hipparcos} satellite, the definition and characterisation of nearby OB associations and open clusters was improved \citep{deZeeuw1999, deBruijne1999, Hoogerwerf1999, Elias2006b,  Elias2006a, Elias2009, Bouy2015}
and our knowledge of the structure of the solar neighbourhood amplified. 

In particular, \cite{Elias2006b} first studied the three dimensional (3D) spatial distribution of early type stars within $1 \, \mathrm{kpc}$ of the Sun by modelling the star distribution with two interacting discs, the Gould Belt  and the Local Galactic Disc.

\cite{Bouy2015} revisited the distribution of OB stars in the solar neighbourhood by constructing a 3D map of their spatial distribution. 
They found three stream-like structures (named Scorpius-Canis Major, Vela, and Orion), not only coherent in space but also characterised by monotonic age sequences. 
The main conclusion emerging from \cite{Elias2006b} and \cite{Bouy2015} is that there is no evidence of a ring-like structure in the 3D configuration of young, bright stars in the solar neighbourhood. 
The Gould Belt as perceived by Herschel and Gould would be due to a projection effect according to Bouy \& Alves (Orion and Sco-Cen causing the apparent tilt due to their locations below and above the plane).

In this work, we make use of the second data release of the \textit{Gaia} mission, \textit{Gaia} DR2, to study the 3D  configuration of the solar neighbourhood, focusing  on young groups and OB associations. We also study the star formation history (SFH) of the solar neighbourhood by estimating the ages of the young groups that we find.

In Sect. \ref{Section2} we give a short description of the data, which we 
divide in two samples, the upper main sequence (UMS) and the pre-main sequence (PMS). We further describe the selection procedure that we used to derive astrometrically `clean' samples, and the photometric and kinematic selection criteria that we apply. In Sect. \ref{Section3} we describe the methods used to obtain a 3D map of the solar neighbourhood, and we study the 3D distribution of the UMS and PMS samples in terms of age. In Sect. \ref{Section4} we discuss our findings. Finally, in Sect. \ref{Section5} we summarise our results and draw our conclusions.

\section{Data}\label{Section2}
In this section we present the selection criteria used for this study. We refer to \cite{Prusti2016, Brown2018} and \cite{Lindegren2018} for a detailed description of the data. The queries that we used to retrieve the data from the \textit{Gaia} archive are reported in Appendix A. 
\newline
We selected all the stars within $d = 500 \, \mathrm{pc}$ of the Sun ($\varpi \ge 2 \, \mathrm{mas}$) and divided them in two samples, the UMS and the PMS. There are two reasons for this division. The first reason concerns the data analysis procedure: dividing the initial sample allows to apply different selection criteria that are more suitable  for one sub-sample or the other. The second reason has instead a scientific justification: it is indeed interesting to study UMS and PMS as two separate samples in order to compare the distribution of young, high-mass stars and low-mass sources.
\newline
Both samples are selected by combining photometric and astrometric criteria. With regards to the photometric criteria, the first step in our procedure consists of correcting for extinction and reddening in the colour-magnitude diagrams. The method that we apply to do such a correction is presented in Section 2.1 
and applied to the UMS and PMS samples in  Sections 2.2 and  2.3, respectively.
The final result of the data selection consists of a catalogue of UMS and PMS stars, which is available on CDS\footnote{The PMS and UMS catalogues are only available in electronic form
at the CDS via anonymous ftp to cdsarc.u-strasbg.fr (130.79.128.5)
or via \url{http://cdsweb.u-strasbg.fr/cgi-bin/qcat?J/A+A/}}.
We shortly describe the catalogue columns in Appendix F.

\subsection{Extinction correction}\label{sec: AG correction}
$G$ band extinction, $A_\mathrm{G}$, and colour excess, $E(G_\mathrm{BP} - G_\mathrm{RP})$, are reported in the \textit{Gaia} DR2 catalogue for a sub-set of sources with measured parallax. Although single extinction and/or reddening values are inaccurate on a star-by-star level, they are mostly unbiased and can be used reliably at the ensemble level \citep{Andrae2018}. 
We can therefore compute extinction (and colour excess) as a function of position and distance, create a 3D $A_\mathrm{G}$ map, and assign to the stars without measured extinction or colour excess a value of $A_\mathrm{G}$ and $E(G_{\mathrm{BP}} - G_{\mathrm{RP}})$ based on the 3D map. In this way, we aim at producing a de-reddened colour-magnitude diagram, to better isolate young star-forming regions.
We use \textit{Gaia} DR2 extinction and reddening values mainly for two reasons. On the one hand, cross-matching with other catalogues such as 2MASS \citep[see e.g.][]{Katz2018, Poggio2018} significantly reduces the number of sources, while we aim to use as many sources as possible. On the other hand, although 3D extinction maps are available, they generally report extinction values in the $V$ band. Thus, one should transfer the $V$ band extinction to the \textit{Gaia} DR2 bands through photometric transformation (or vice-versa). Even though this is in principle possible, it is very error-prone as the transformation between $A_\mathrm{V}$ and $A_\mathrm{G}$ and between $E(B-V)$ and $E(G_{\mathrm{BP}} - G_{\mathrm{RP}})$ is non-trivial due to the very wide photometric bands used by \textit{Gaia} (see \cite{Andrae2018} for more details).
\newline 
To create the map, we proceed as follows. 
We query all the sources with $\varpi > 2 \, \mathrm{mas}$,  $\varpi/ \sigma_\varpi > 5 $ and a measured $A_\mathrm{G}$ value. 
We then compute the source galactic Cartesian coordinates, $x, y, z$.
We define a volume $N = 1000 \times 1000 \times 1000 \, \mathrm{pc}$ centred on the Sun and we divide it into cubes $n$ of 
$10 \times 10 \times 10 \, \mathrm{pc}$ 
each.  For each cube, we compute the average extinction and colour excess.
In this way, we obtain a crude map that nonetheless delivers better results than the alternatives described above. 
Finally, we assign to all the sources the appropriate extinction and colour excess values according to their position in space, and we correct the observed $M_G$ versus $G_{\mathrm{BP}} - G_{\mathrm{RP}}$ colour-magnitude diagram.

\subsection{Upper main sequence}\label{sec:UMS}
\begin{figure}
\includegraphics[width=\hsize]{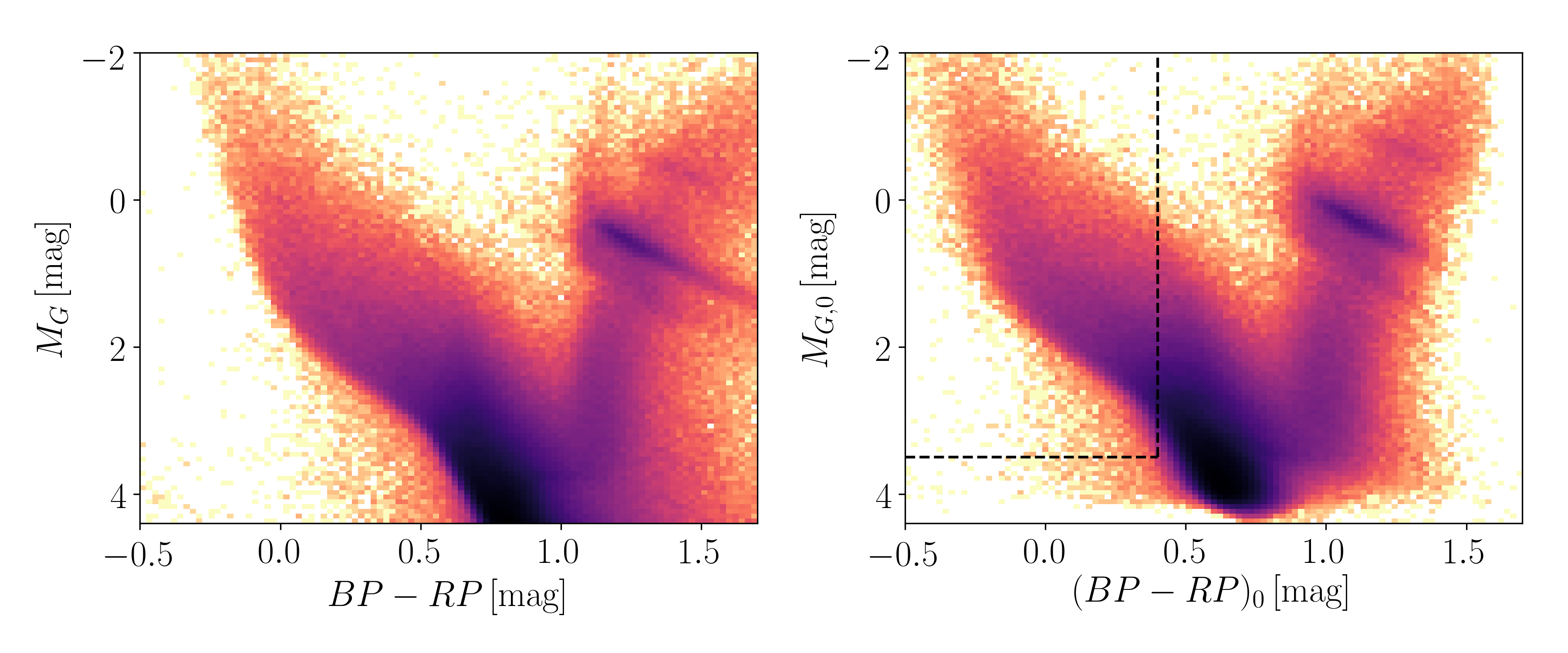}
\caption{UMS colour-magnitude diagrams. Left: Colour-magnitude diagram before correcting for extinction and colour excess. Right: Colour-magnitude diagram after accounting for extinction and reddening.
The dashed lines limit the region we considered as the UMS in this study.}
\label{fig:fig111}
\end{figure}

To construct the sample, we first downloaded from the \textit{Gaia} archive
bright and blue sources nominally closer than $d = 500 \, \mathrm{pc}$ to the Sun:
\begin{align}
M_{G} &\le 4.4 \, \mathrm{mag}; \nonumber \\
(G_{\mathrm{BP}} - G_{\mathrm{RP}}) & \le 1.7\, \mathrm{mag}; \nonumber\\ 
\varpi & >= 2 \, \mathrm{mas};\\
\varpi/\sigma_{\varpi} & > 5.
\end{align}
By using the extinction $A_G$ and colour excess $E(G_{BP} - G_{RP})$ values computed in Section \ref{sec: AG correction}, we correct the colour-magnitude diagram for extinction and reddening, and apply the following selection criteria:
\begin{align}\label{eq:eq2}
M_{G,0} &\le 3.5 \, \mathrm{mag}, \nonumber \\
(G_{\mathrm{BP}} - G_{\mathrm{RP}})_0 & \le 0.4 \, \mathrm{mag}. \nonumber\\ 
\end{align}
The first and second condition aim at selecting sources whose colours are consistent with being of spectral type O, B, or A. The condition $\varpi/\sigma_{\varpi} > 5$ is primarily motivated by  the fact that in the rest of the paper we compute distances simply by inverting parallaxes, ($d = 1000/\varpi  \, \mathrm{pc}$), and this holds only when parallax errors are small  \citep{BailerJones2015}.
Figure \ref{fig:fig111}(left) shows the initial colour-magnitude diagram used for the selection. Figure \ref{fig:fig111}(right) shows the conditions on colour and magnitude as black dashed lines.

\begin{figure*}
\includegraphics[width=\hsize]{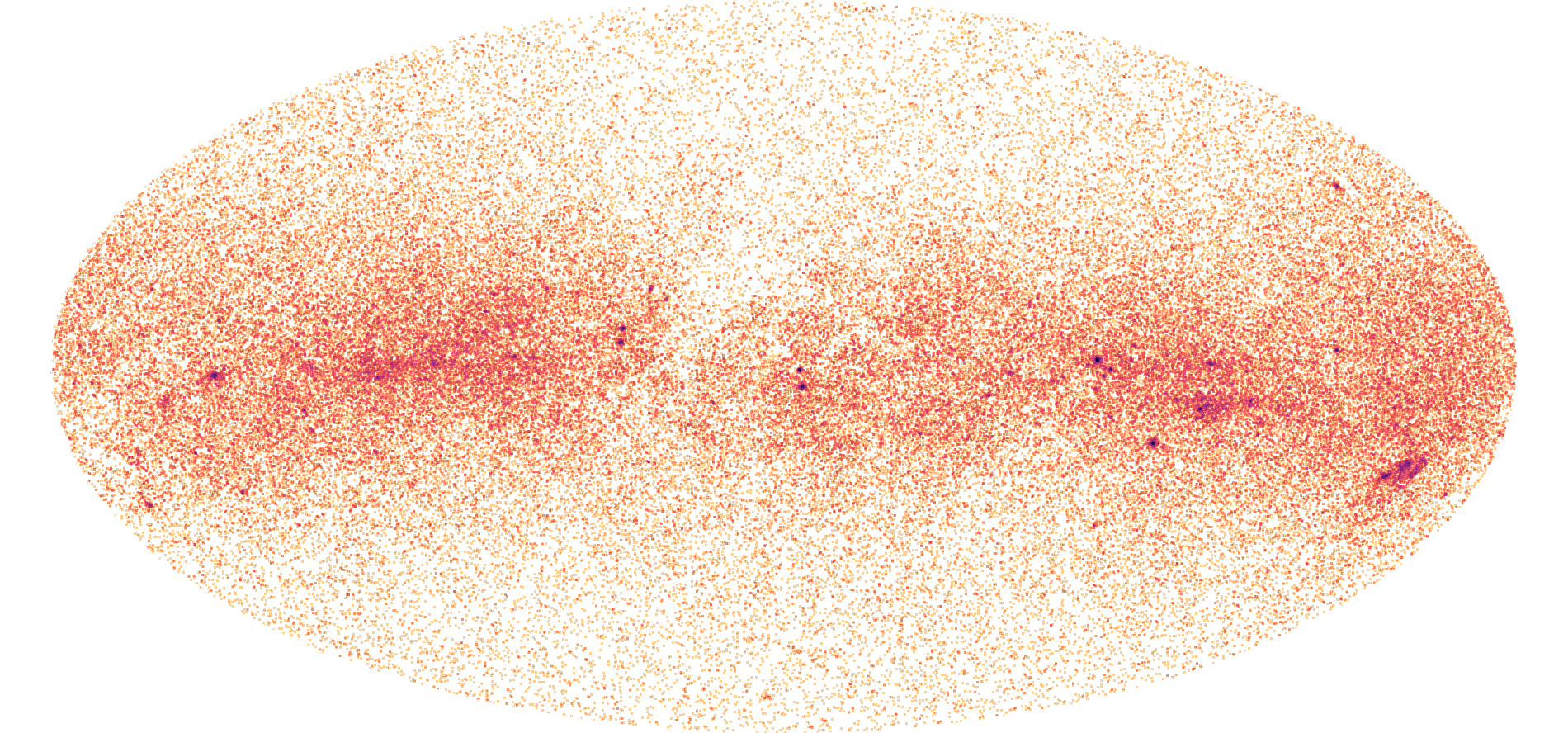}
\caption{UMS sources selected by applying the conditions detailed in Sect. \ref{sec:UMS}. The sources are concentrated towards the galactic plane,
and their density decreases towards the poles. Clumps corresponding to known open clusters and associations are visible.}
\label{fig:fig1}
\end{figure*}

\subsubsection{Tangential velocities}\label{sec:UMS vt}
\begin{figure}
\includegraphics[width=\hsize]{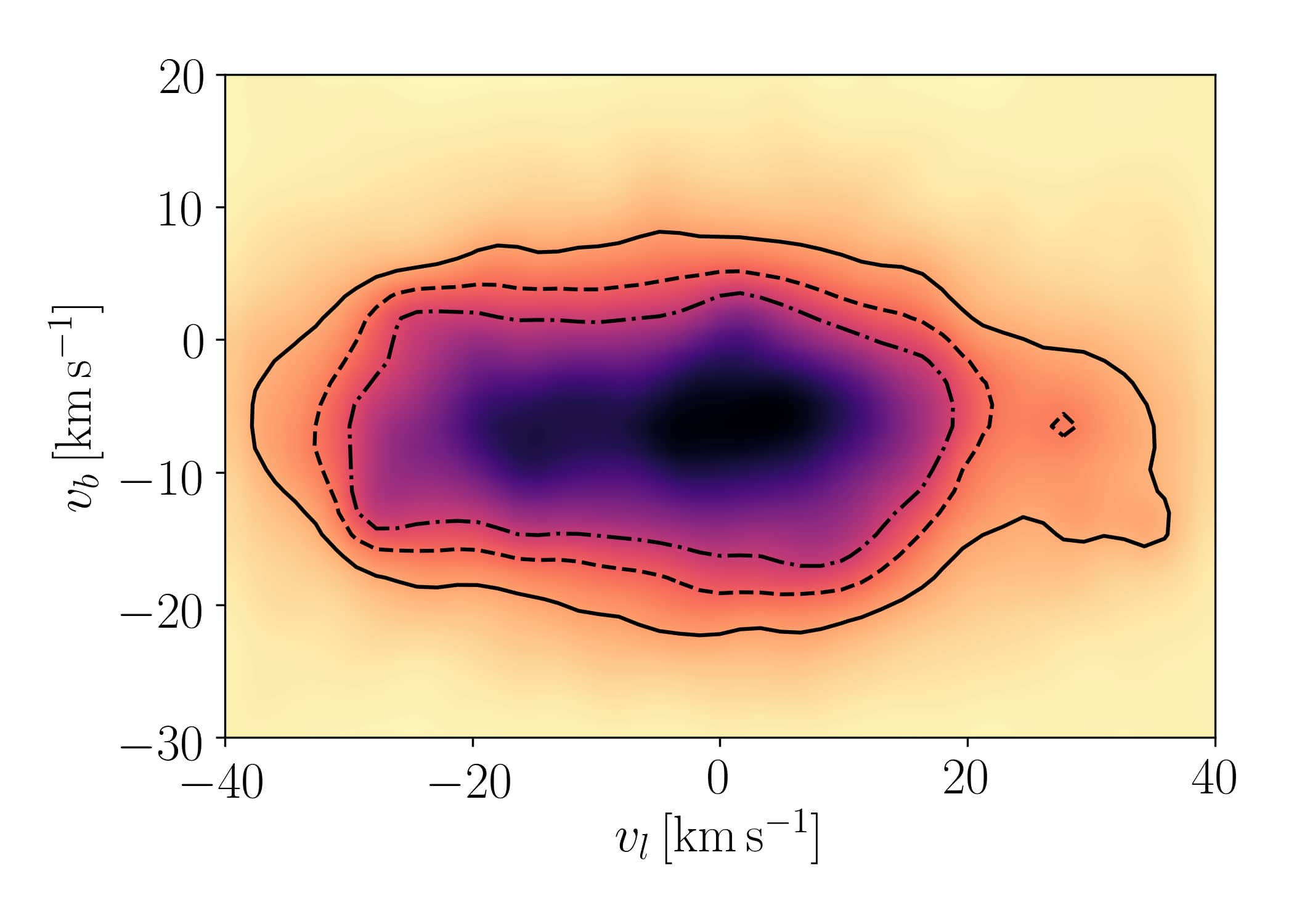}
\caption{Smoothed tangential velocity distribution of the UMS sample, defined in Eq. 3 in the text. The contours represent the $S = 1, 2, 3$ levels. The density enhancements correspond to known clusters and associations. We also note that the distribution is not centred in $v_l, v_b = (0, 0)$ due to the solar motion.}
\label{fig:fig211}
\end{figure}

\begin{figure*}
\includegraphics[width=\hsize]{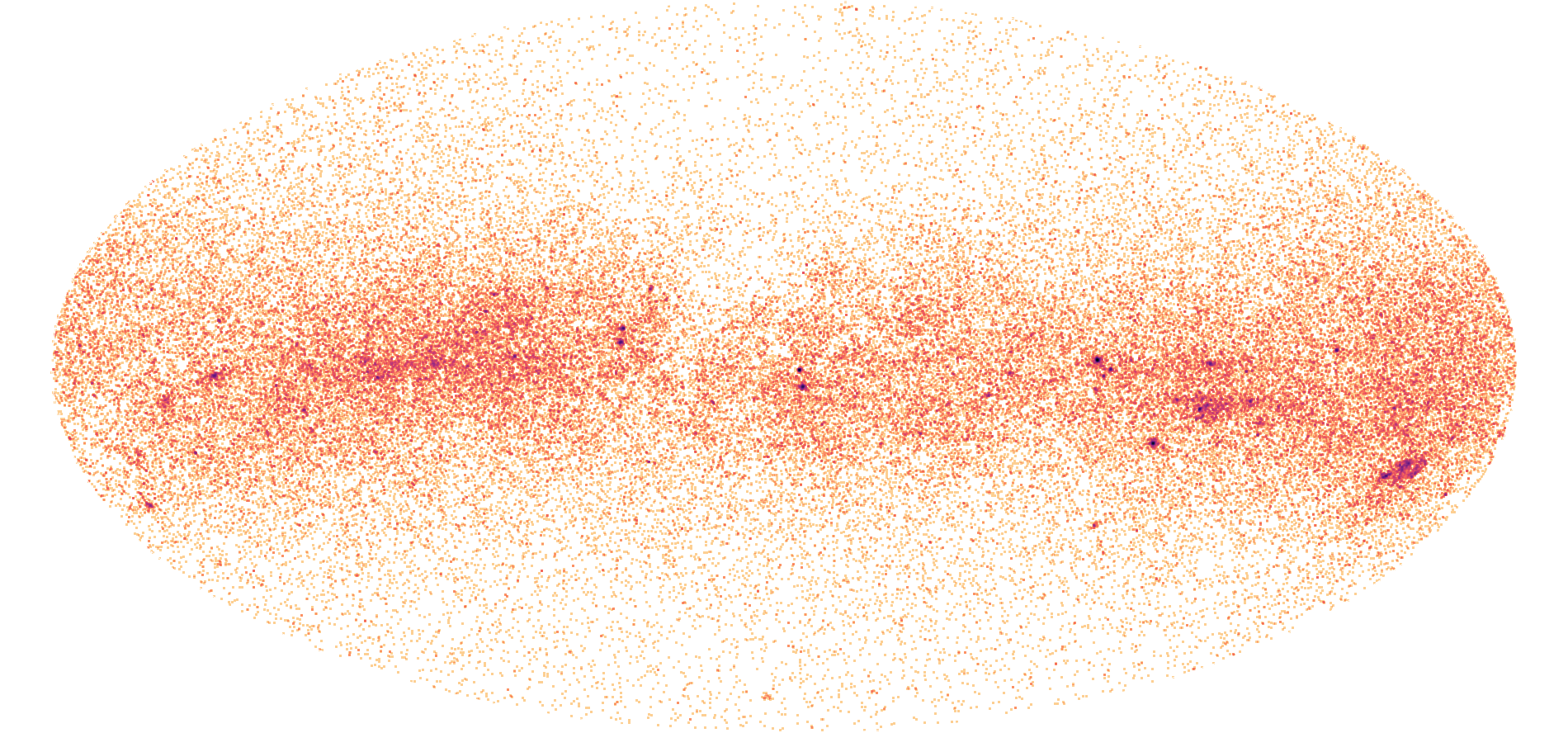}
\caption{Distribution of the sources in the Sky after the selection based on tangential velocities. The number of sources at high galactic latitudes has decreased with respect to Fig. \ref{fig:fig1}, which indicates that many contaminants have been discarded. }
\label{fig:fig2112}
\end{figure*}

Figure \ref{fig:fig1} shows the distribution of the UMS sources selected in Sect. \ref{sec:UMS}. The density of sources increases towards the galactic plane, and some known clusters are  visible. 
Members of clusters and associations share the same spatial velocity, with a small velocity dispersion that varies from a few hundred metres per second to some kilometres per second, respectively. In proper motion or tangential velocity space, they appear as density enhancements with respect to the underlying, broad  
field star distribution. 
Therefore, to clean our sample, we study the tangential velocity distribution
($v_{l, b} = A \mu_{l*,b} / \varpi$, where $A = 4.74047 \, \mathrm{km s^{-1} yr^{-1}} $) of the stars we have selected so far. 
\newline
Figure \ref{fig:fig211} shows an unsharp mask of the tangential velocity distribution of the UMS sample. We use a two-dimensional (2D) Gaussian filter with bandwidth $= 30 \, \mathrm{km \, s^{-1}}$ to smooth the tangential velocity distribution. This produces a blurred (`unsharp') mask of the original distribution. The unsharp mask is subtracted from the original tangential velocity distribution, which was smoothed as well with a Gaussian filter of bandwidth $= 1 \, \mathrm{km \, s^{-1}} $. Finally we compute the quantity
\begin{equation}
S = \frac{I_{1} - I_{30}}{I_{30}},
\end{equation}
where $I_{x}$ represents the smoothed tangential velocity distribution and $S$ is then a measure of the contrast of the density enhancements with respect to a uniform, smooth distribution. We selected the stars within the $S = 1$ levels, shown as black solid lines in Fig. \ref{fig:fig211}.
Figure \ref{fig:fig2112} shows the distribution in the sky of the sources selected in this fashion. The number of sources at high galactic latitudes visibly decreases with respect to Fig. \ref{fig:fig1}, indicating that the tangential velocity selection is useful to reduce the contamination level of our sample, since we expect young stars to be mainly located towards the galactic plane. On the other hand, such a selection will reject young stars with peculiar tangential velocities (such as binaries or runaways): we stress however that the focus of this study is on the bulk of the early-type population and not on the kinematic outliers, which represent a small fraction of the population.

\subsection{Pre-main sequence}\label{Subsec:PMS}
To select the PMS sample, we first downloaded from the \textit{Gaia} archive all the sources nominally within $d = 500 \, \mathrm{pc}$. Due to the large number of sources, the query cannot be executed as a single query, but the data has to be divided, for example in parallax bins.
After joining all the separate tables, we proceed as follows.

\noindent
\subsubsection{Astrometrically `clean' subset}
We first applied Eqs. C.1 and C.2 of \cite{Lindegren2018} and required that $\varpi/\sigma_{\varpi} > 5$. Equations C.1 and C.2 were used by \cite{Lindegren2018} to produce a 'clean' HR diagram of nearby stars ($d < 100 \, \mathrm{pc}$). Equation C.1 is meant to remove sources with spuriously high parallax.
Equation C.2 deals with the photometric errors in the BP and RP bands, affecting faint sources and crowded areas
in particular. We selected stars with small parallax error ($\sigma_{\varpi}/\varpi < 20 \%$) with the same motivations as for the UMS sample.
Finally, we decided to restrict our sample to stars following the disc kinematics. Thus we required the total tangential velocity to be lower than $40 \, \mathrm{km \, s^{-1}}$: \\
$v_t = \sqrt{v_{l}^2 + v_{b}^2} < 40 \, \mathrm{km \, s^{-1}} $. \\ 
\noindent
The condition on the tangential velocity follows \cite{Babusiaux2018}. Usually the cut to select thin disc stars is $v_{\mathrm{TOT}} < 50 \, \mathrm{km \, s^{-1}}$ \citep[e.g.][]{Bensby2014}, however we only have two velocity components instead of three, and therefore we adapted the cut to take this into account.

\subsubsection{Extinction correction and selection of the PMS}
\begin{figure}
\includegraphics[width=\hsize]{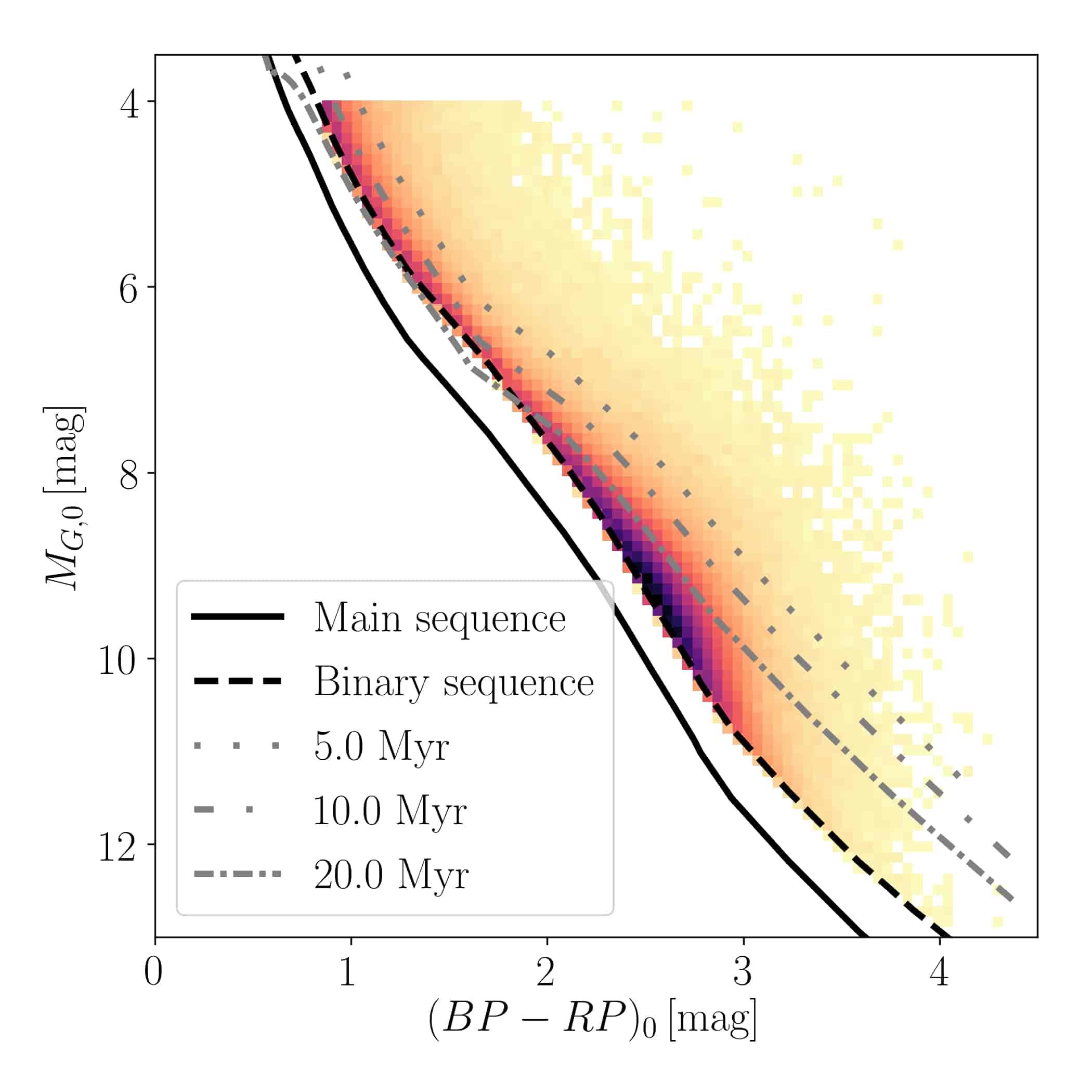}
\caption{$G_{\mathrm{BP}} - G_{\mathrm{RP}}$ vs. $M_G$ colour-magnitude diagram of the sources selected in Sect. 2.2.2. The density of sources increases towards the binary sequence. }
\label{fig:fig12}
\end{figure}

\begin{figure*}
\begin{center}
\includegraphics[width=0.9\hsize]{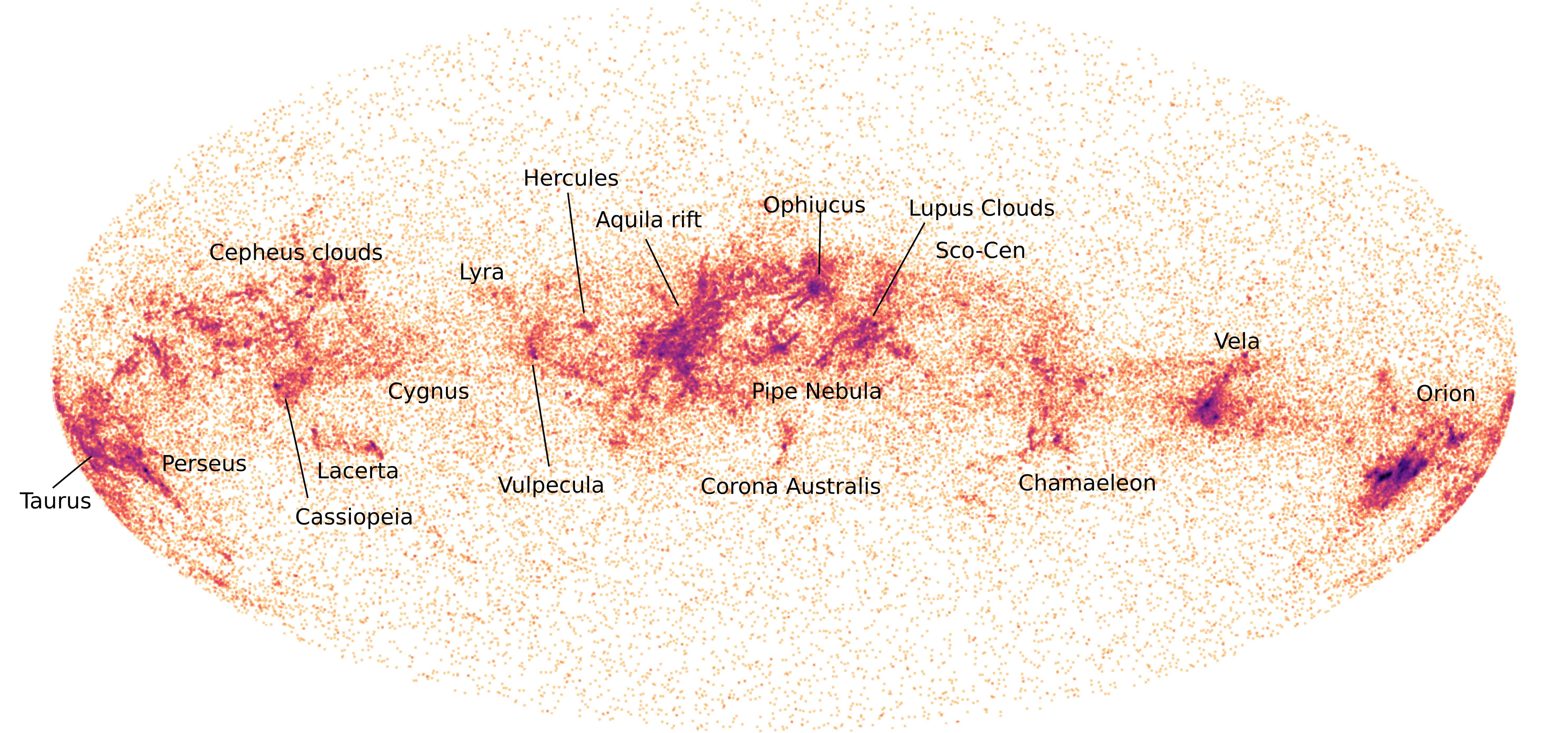}
\includegraphics[width=0.9\hsize]{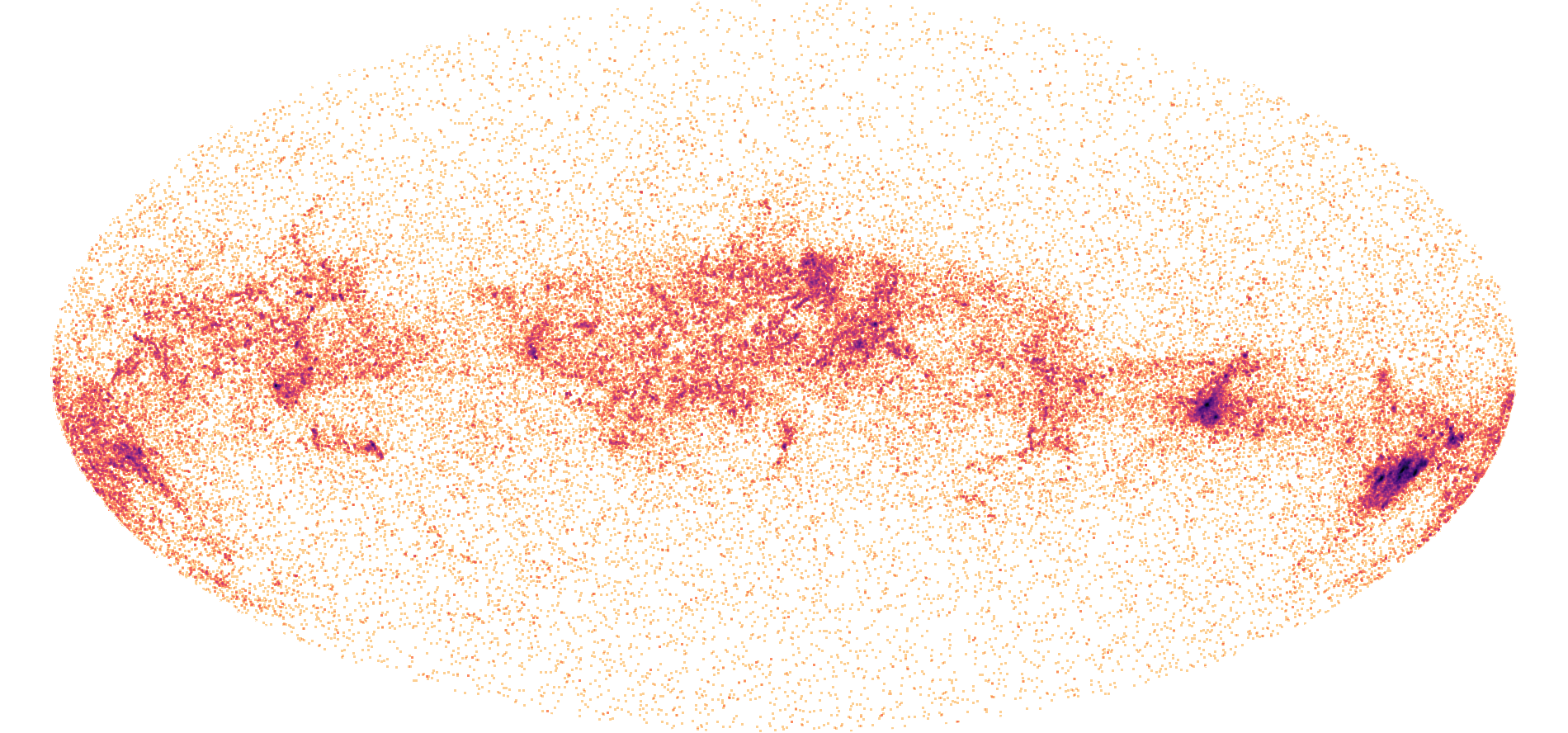}
\includegraphics[width=0.9\hsize]{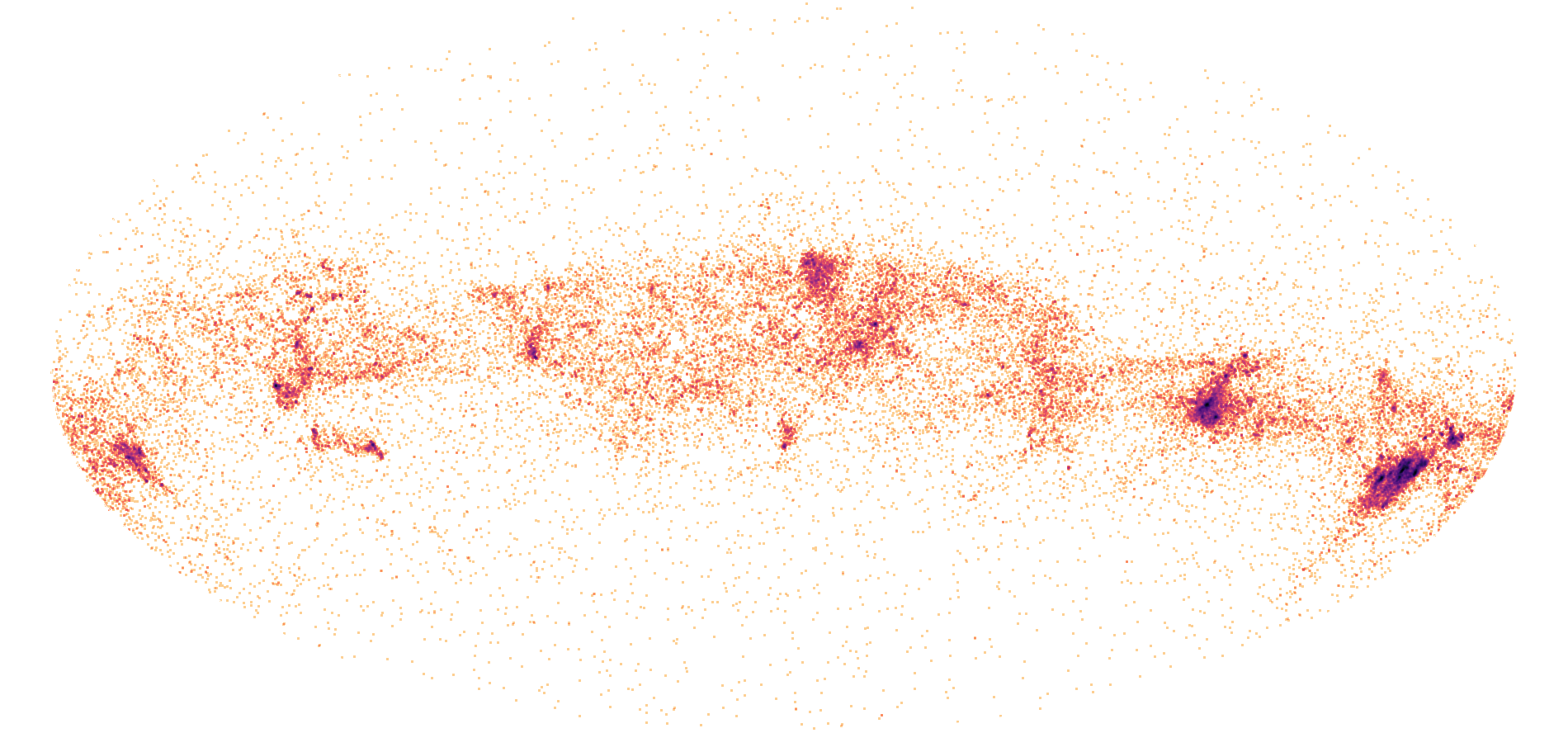}
\end{center}
\caption{Top: PMS sources younger than 20 Myr. Centre: PMS sources younger than 20 Myr, with $A_\mathrm{G} < 0.92 \, \mathrm{mag}$. Bottom:  PMS sources younger than 20 Myr, 
with $A_\mathrm{G} < 0.92 \, \mathrm{mag}$, and within the $S =1$ level of Fig. \ref{fig:fig3}.}
\label{fig:fig15}
\end{figure*}

We first corrected for extinction and reddening using the procedure described in Section \ref{sec: AG correction}.
Then, we used the PARSEC Isochrones \citep{Bressan2012} version 1.2S \citep{Chen2014, Chen2015, Tang2014} with $A_V = 0 \, \mathrm{mag}$ and solar metallicity ($Z = 0.0152$) to define the MS track and the binary sequence (which is brighter than the MS by 0.75 mag); finally, we selected all the stars brighter than the binary sequence. We further restrict our sample to sources with $M_{G,0} > 4 \, \mathrm{mag}$:  this cut is motivated by the need to exclude sources that are located on the MS turn-off and on the faint end of the giant branch. Figure \ref{fig:fig12} shows the color magnitude diagram of the selection.  We note that for $M_{G,0}\sim 7 \, \mathrm{mag}$ the binary sequence (black dashed line) and the 20 Myr isochrone (grey dotted line) overlap; 
therefore we expect that region of the colour-magnitude diagram to be contaminated by old binaries (see Sect. 3.4 for a more detailed discussion). In general, the area of the colour-magnitude diagram next to the binary sequence is bound to be subject to contamination from unresolved binaries, but also from reddened MS sources: to partially eliminate the issue, we decided to restrict our sample further, to the sources brighter (and therefore younger) than the 20 Myr isochrone \footnote{We also tested whether we would obtain different results by considering, for instance, the luminosity above the MS as an age proxy: this was not the case.}. Figure \ref{fig:fig15} (top) shows the position in the sky of the sources selected with this procedure.  Some groups can be easily identified:
\begin{itemize}
\item \textit{Orion}, on the rightmost side at $l < 220^{\circ}$;
\item \textit{Vela}, at $240^{\circ} < l < 270^{\circ}$;
\item  \textit{Scorpius-Centaurus} and \textit{Ophiucus}, at $l > 280^{\circ}$ and positive $b$;
\item \textit{Chamaeleon}, at $l, b \sim (300^{\circ}, -16^{\circ})$;
\item The \textit{Aquila rift}, at $l, b \sim (30^{\circ}, +3^{\circ})$;
\item \textit{Lacerta}, at $\sim (100^{\circ}, -20^{\circ})$;
\item \textit{Cepheus} and \textit{Cassiopeia}, at $l > 100^{\circ}$, above and slightly below the galactic plane;
\item \textit{Taurus} and \textit{Perseus}, at $l > 140^{\circ}$, below the galactic plane.
\end{itemize}
The source distribution follows the dust features located in the galactic plane: while on the one hand it is expected that young sources follow the outline of the molecular clouds, on the other hand it is likely that our sample is still contaminated by MS stars located behind the molecular clouds.  Therefore, to remove the last contaminants, we discarded all the sources with $A_\mathrm{G} > 0.92 \, \mathrm{mag}$. We chose this threshold after studying the extinction distribution of our sample: the median of the distribution is $0.51 \, \mathrm{mag}$, while the $16th$ percentile is $0.30 \, \mathrm{mag}$ and the $84th$ percentile is $0.92 \, \mathrm{mag}$. Thus, we excluded all the sources with extinction larger than the $84th$ percentile. This is a rough cut that might exclude not only reddened MS sources, but also young sources embedded in the clouds; however it is on average effective in removing contaminants (see also Appendix E). Figure \ref{fig:fig15} (centre) shows the distribution in the sky of the sources remaining after the extinction cut.

\subsubsection{Tangential velocities}
\begin{figure}
\includegraphics[width=\hsize]{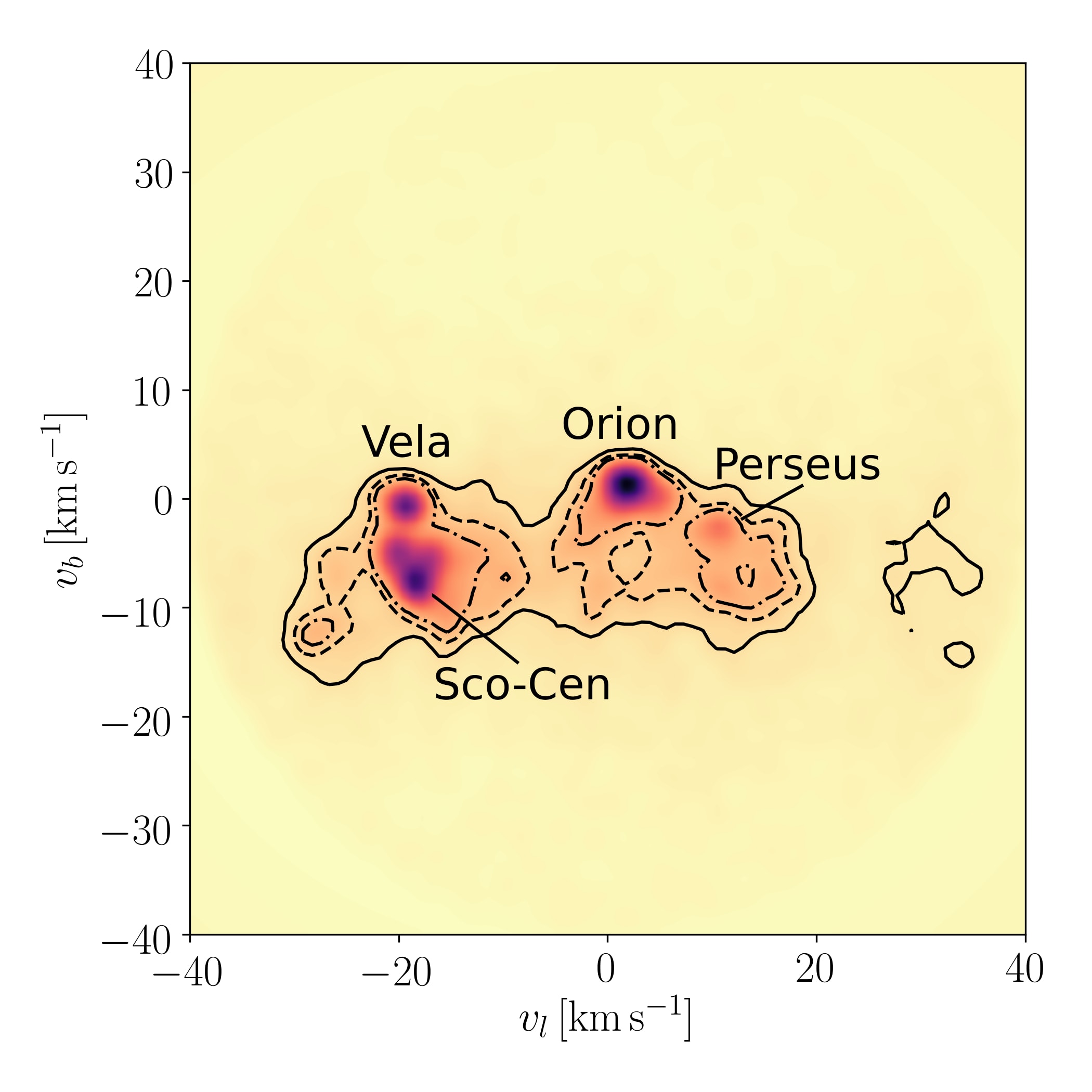}
\caption{Tangential velocity plot of the PMS sample selected in Section 2.
Many clumps are visible and correspond to known associations and clusters. The four most prominent structures are: Orion, Sco-Cen, Vela, and Perseus. We note the gap around $20 \, \mathrm{km \, s^{-1}}$, visible also in Fig. \ref{fig:fig24}.}
\label{fig:fig3}
\end{figure}

As in Sect. \ref{sec:UMS vt}, we finally perform a selection in tangential velocity space, relying on the fact that the young clusters and associations that we are interested in share the same kinematic properties.
Figure \ref{fig:fig3} shows the tangential velocity distribution defined in Eq. \ref{eq:eq2} of the sources selected in Sect. 2.2.2.
The contour lines represent the $S = 1, 2, 3$ levels.
Analogously as with the UMS sample, we selected all the sources within the $S = 1$ level. 
The final PMS sample is shown at the bottom of Fig. \ref{fig:fig15}. 
As mentioned in the previous section, the extinction correction reduces the imprint of the molecular clouds on the star distribution. The tangential velocity selection instead mostly reduces the number of sources at high galactic latitudes.

\section{Three dimensional mapping of young stars in the solar neighbourhood}
In this section we describe the method we use to make 3D 
density maps of the solar neighbourhood. 
We make two maps, one for the UMS sample and one for the PMS sample. The maps are then discussed and compared in this section and in Sect. \ref{Section4}.
\label{Section3}
\subsection{Method}
Similarly to what we did in Sect. 2.2.3, the first step of creating the maps is to compute galactic Cartesian coordinates, $x, y, z$,  
for all the sources and to define a box $V$ = $1000 \times 1000 \times 700 \,  \mathrm{pc}$ centred on the Sun. We divide the cube into volume elements $v$  of  $3\times 3\times 3 \,  \mathrm{pc}$. After computing the number of stars in each volume $n$, we estimate the star density $D(x, y, z)$ by smoothing the distribution by means of a 3D Gaussian filter, using a  technique similar to that used by \cite{Bouy2015}.

The Gaussian width (equal on the three axes) is $w = 3$ pc for PMS stars and $w = 4$ pc for UMS stars, and the Gaussian is  truncated at $3\sigma$ (\footnote{The python function used for the smoothing is scipy.ndimage.filters.gaussian\_filter()}).
The choice of a certain $w$ value is arbitrary. A  high $w$ value produces a smooth, less detailed map, while a low $w$ value results in a noisy map.  
We finally normalize the density distribution by applying the  \textit{sigmoidal logistic} function:
\begin{equation}
f(\mathbf{x}) = \frac{L}{1 + e^{-k(\mathbf{x}-\mathbf{x_0})}} - 1,  
\end{equation}
where $\mathbf{x} = D(x, y, z)$, and $D$ is the not normalized density distribution. The parameters we chose are: $L = 2, x_0 = 0 , k = 30$ for PMS stars; and $L = 2, x_0 = 0 , k = 40$ for UMS stars. In this way, $f(\mathbf{x})$ ranges from 0 to 1 as $x$ ranges between 0 and infinity.
A low $k$ value reveals more detail at higher densities and a high $k$ value reveals more detail at lower densities.
The choice of the appropriate Gaussian $w$ value and logistic $k$ value depends upon the desired map presentation. We have chosen the best values to visualise  stellar concentrations for the UMS and PMS maps.

\subsection{Results}
\begin{figure*}
\begin{minipage}{0.5\hsize}
\includegraphics[width=\hsize]{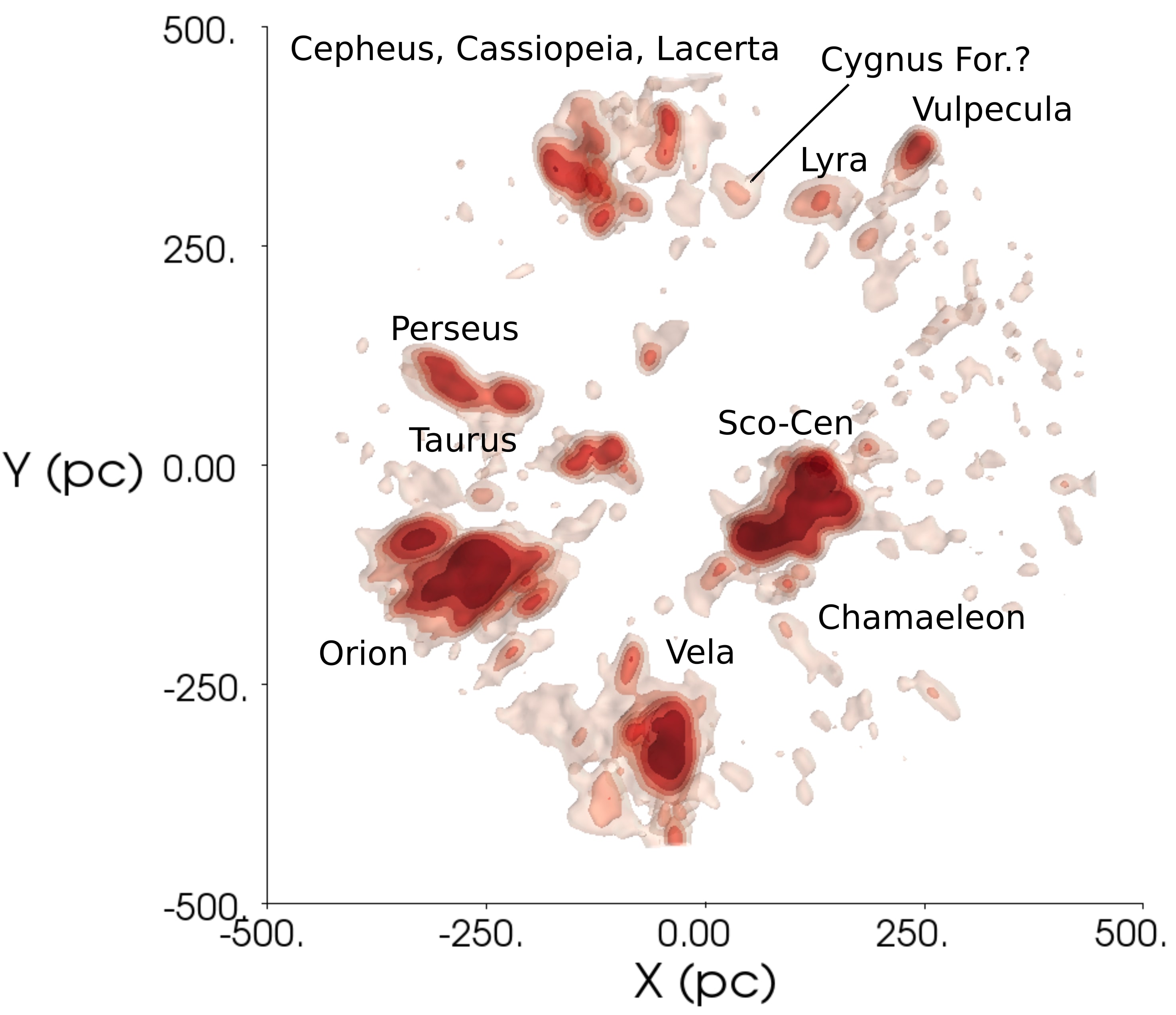}
\end{minipage}
\begin{minipage}{0.5\hsize}
\includegraphics[width=\hsize, height = 0.2\vsize]{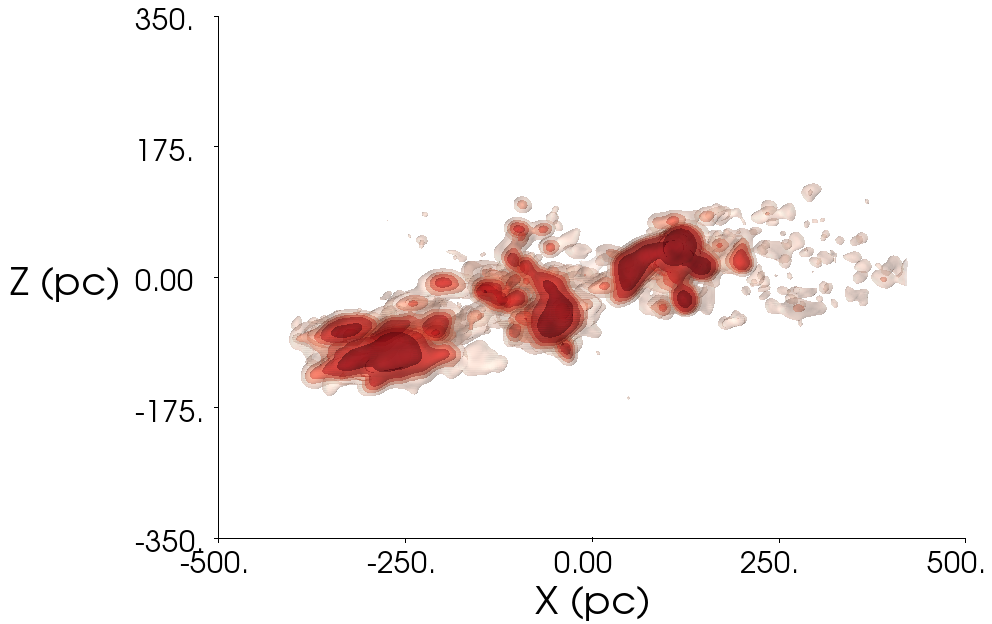}\\
\includegraphics[width=\hsize, height = 0.2\vsize]{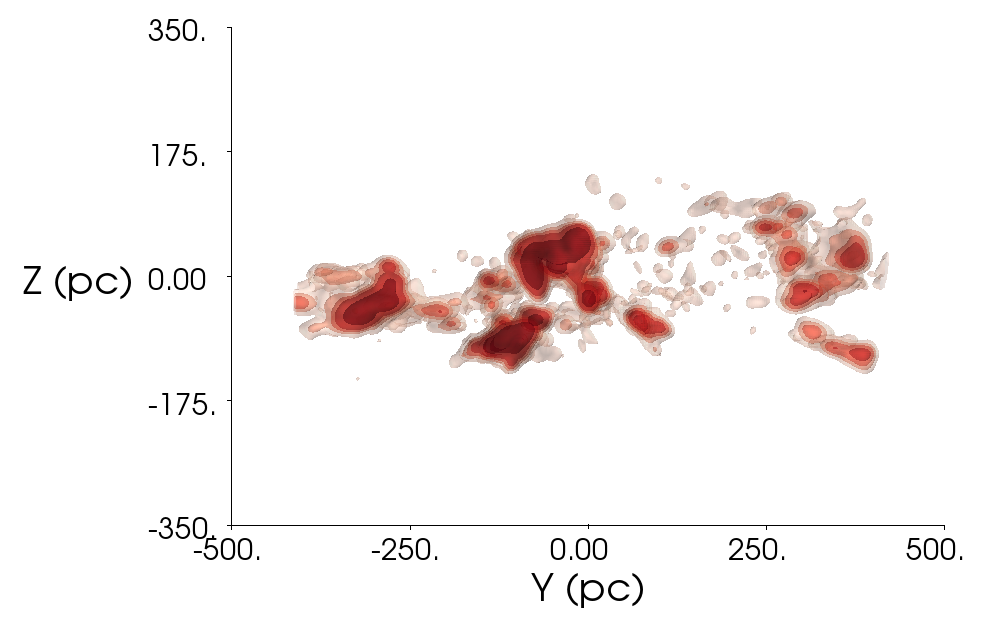}
\end{minipage}
\caption{\textit{Left}: 3D density distribution of PMS sources younger than 20 Myr on the galactic plane. The Sun is in (0, 0), the x-axis is directed towards the galactic centre, and the y-axis towards the direction of the galactic rotation. The z-axis is perpendicular to the plane. The contours represent the  0.2, 0.4, 0.6, 0.8, and 1 density levels. Right, top: 3D density distribution of the PMS sample (age < 20 Myr) perpendicular to the galactic plane. Contour levels are the same as on the left. \textit{Right, bottom}: 3D density distribution of the PMS sample (age < 20 Myr) along the rotation axis. 
}
\label{fig:fig2}
\end{figure*}

\begin{figure*}
\begin{minipage}{0.5\hsize}
\includegraphics[width=\hsize]{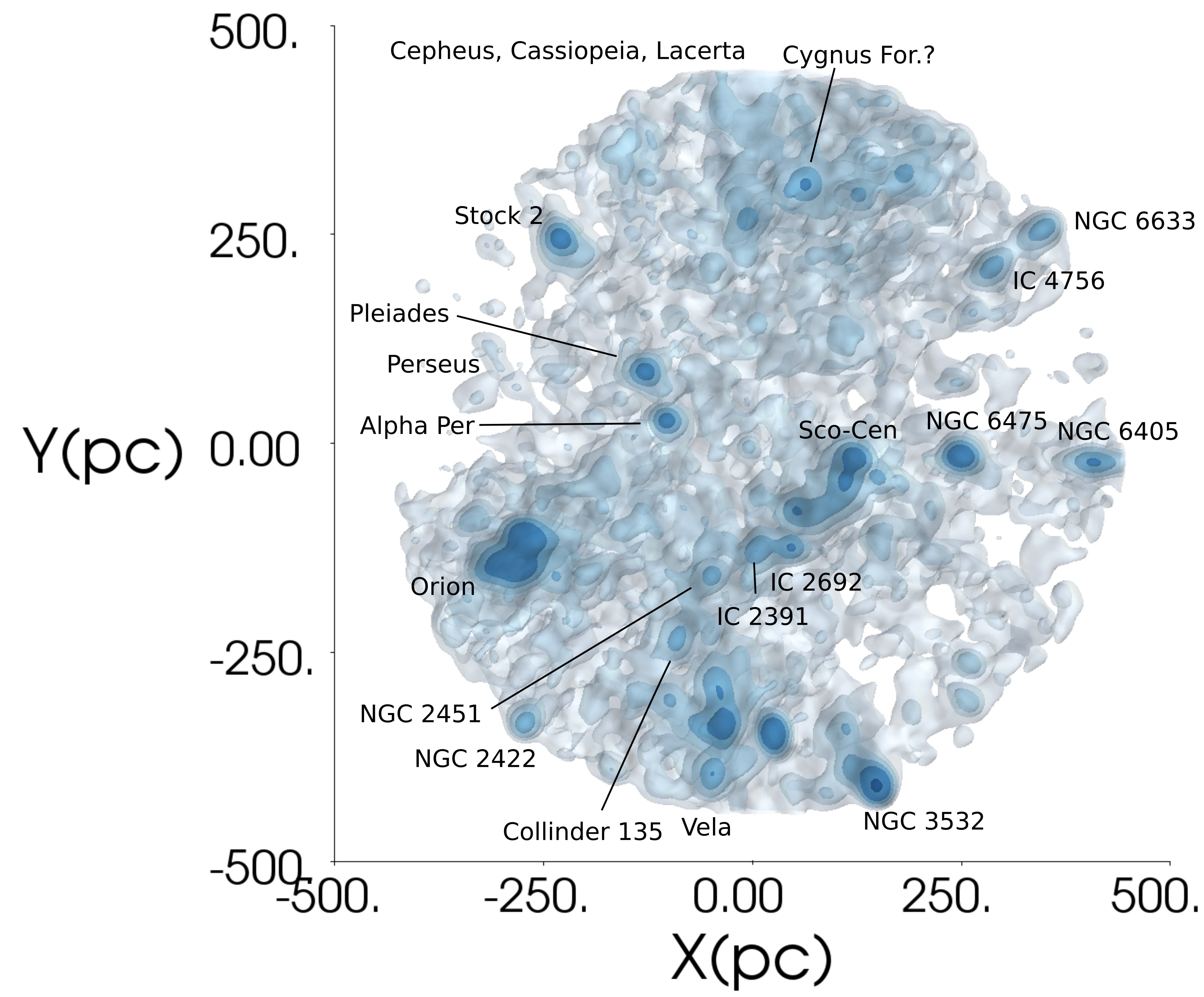}
\end{minipage}
\begin{minipage}{0.5\hsize}%
\includegraphics[width=\hsize, height = 0.2\vsize]{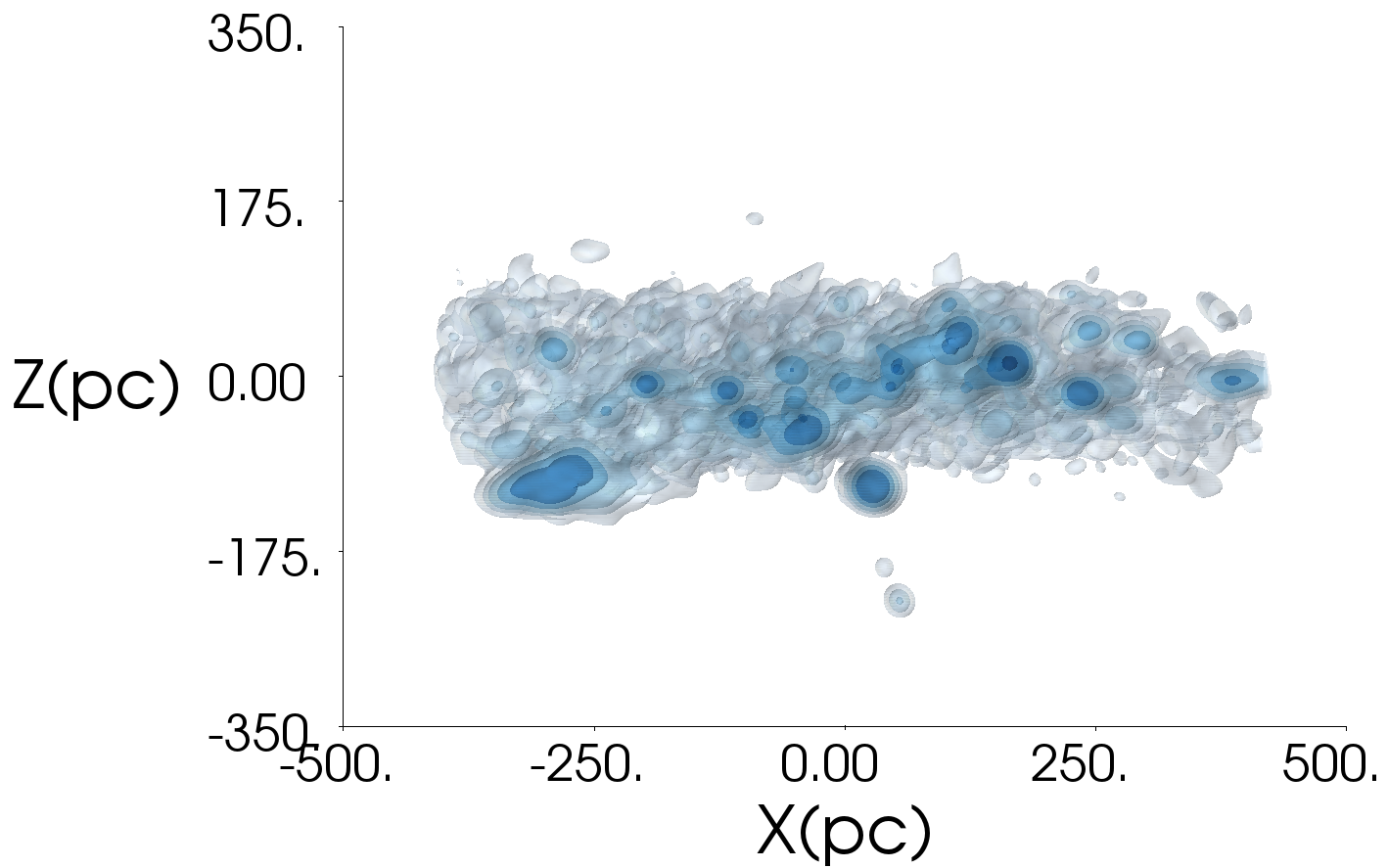}\\
\includegraphics[width=\hsize, height = 0.2\vsize]{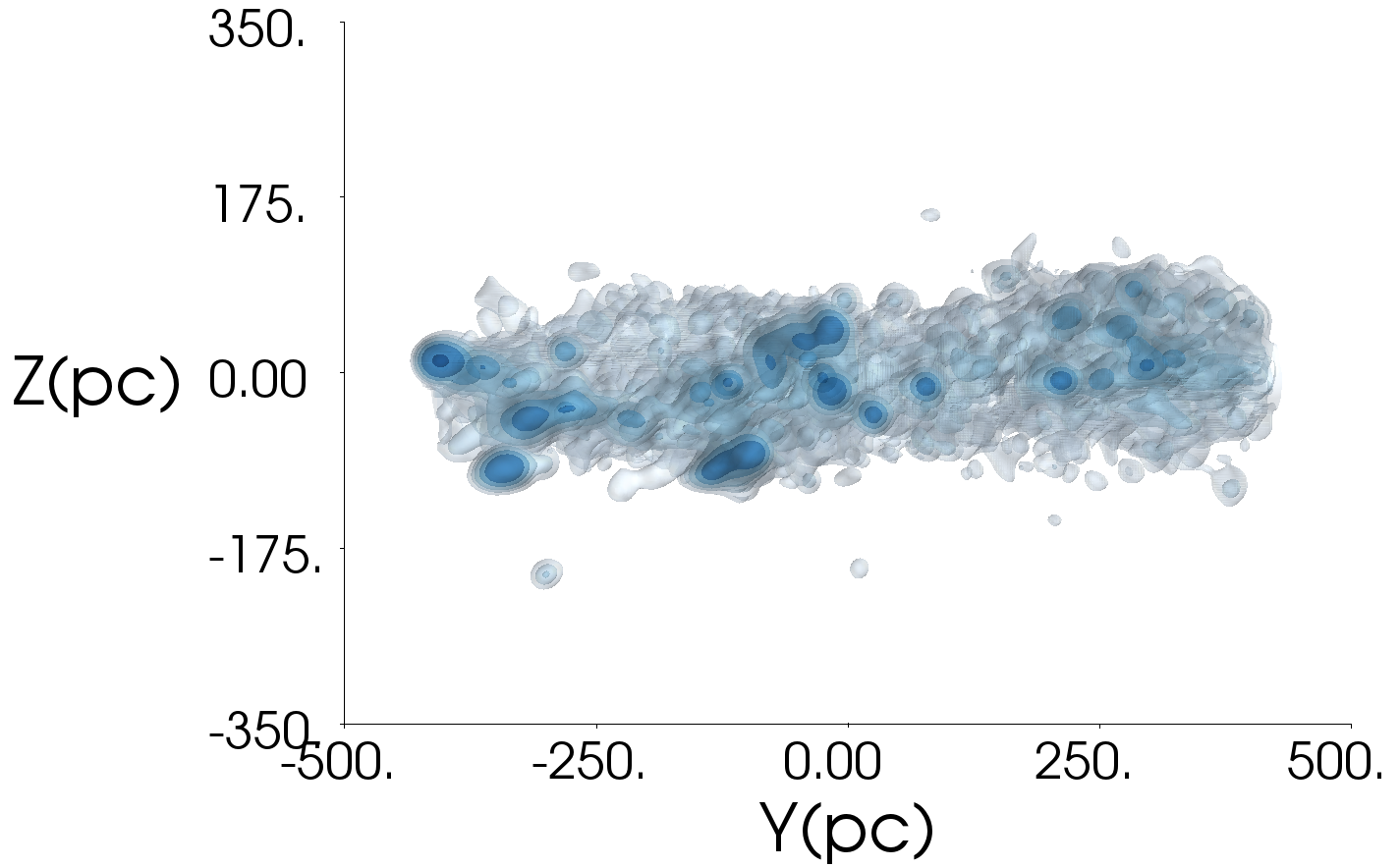}
\end{minipage}
\caption{As in Fig. \ref{fig:fig2}, but for the UMS sample selected in Section 2.1.  The contours represent the  0.2, 0.3, 0.4, 0.6, 0.8, and 1 density levels.}
\label{fig:fig21}
\end{figure*}

Figure \ref{fig:fig2} (left) shows the density distribution of PMS sources younger than $ 20 \, \mathrm{Myr}$ on the galactic plane ($X$ is directed towards the galactic centre, $Y$ towards galactic rotation, and the Sun is at $(0, 0, 0)$). Figure \ref{fig:fig2} (right) shows the density distribution perpendicular to the plane.
Figure  \ref{fig:fig21} shows the density distribution of the UMS sample.  The axes are the same as in Fig. \ref{fig:fig2}. 

\noindent
Three main density enhancements visible in both maps are the following.
\begin{enumerate}
\item \textit{Scorpius-Centaurus (Sco OB2)}: $0 < X < 250$ pc and $-200 < Y <  0 $ pc.\\
Due to its proximity \citep[$d\sim 140 \, \mathrm{pc}$,][]{deZeeuw1999}, the Sco OB2 has been extensively studied \citep{deBruijne1999b, Rizzuto2011, Pecaut2012, Wright2018}.
The association is usually divided into three subgroups, \textit{Upper Scorpius} (US), \textit{Upper Centaurus-Lupus} (UCL), and \textit{Lower Centaurus-Crux} (LCC), with median ages of 11, 16, and 17 Myr, respectively \citep{Pecaut2016}.  
\item \textit{Vela (Vel OB2)}: $-100 < X < 100 $ and $-100 < Y < -450$ pc.

Vel OB2 has a distance of $d \sim \, 410 \, \mathrm{pc}$.
\cite{Sacco2015}, \cite{Jeffries2014}, \cite{Damiani2017}, and \cite{Franciosini2018} studied the stellar population towards the Gamma Vel cluster and NGC 2547, finding kinematically distinct populations. Using \textit{Gaia} DR1 and \textit{Gaia} DR2, respectively, \cite{Armstrong2018} and \cite{Beccari2018} recently found that the association is composed of many young clusters. In particular \cite{Beccari2018} discovered four new clusters, in addition to Gamma Vel and NGC 2547; four of these clusters are coeval and formed $\sim 10 \, \mathrm{Myr}$ ago, while NGC 2547 and a newly discovered cluster formed $\sim 30 \, \mathrm{Myr}$ ago.  \cite{Cantat2018} also characterised the distribution of  Vel OB2 on a large spatial scale, and found that the distribution of young stars traces the IRAS Vela Shell. This might suggest a common history for Vel OB2 and the Vela Shell: a previous star formation event 
caused the expansion of the shell and likely triggered the formation of the clusters composing the association.
\item \textit{Orion (Ori OB1)}: $-300 < X < -200$ and $-200 < Y < -100 $ pc.

Orion is the nearest ($d \sim 400 \, \mathrm{pc}$) giant molecular cloud complex and it is a site of active star formation, including high-mass stars \citep[e.g.][ and references therein]{Bally2008}. \cite{Zari2017} used \textit{Gaia} DR1 to explore the arrangement and the age ordering of the numerous stellar groups towards the Orion OB association. \cite{Kounkel2018} used \textit{Gaia} DR2 and APOGEE-2 
to identify spatially and kinematically distinct young groups.
\end{enumerate}
The PMS population of Sco OB2, Vel OB2, and Ori OB1 is predominantly concentrated in the dense areas of the UMS population. The latter appears, instead, more diffuse, almost connecting the three regions. A few, more evolved clusters are also visible in Fig. \ref{fig:fig21}: IC 2602, IC 2391, NGC 2451, NGC 2516, NGC 3532, NGC 2422,  NGC 6475, NGC 6405, IC 4756, NGC 6633, NGC 7092, Stock 2, $\alpha$ Per, and Pleiades. Some of these clusters appear embedded in the low-density levels of the UMS density distribution: this might suggest a relation between current star-forming regions and previous star-formation episodes. Finally, it is particularly interesting to notice the presence of a diffuse population in front of the Orion complex (visible in both the UMS map of Fig. \ref{fig:fig21} and the PMS map of Fig. \ref{fig:fig2}). This population was already observed by \cite{Bouy2015, Zari2017} and \cite{Kounkel2018}, and here we confirm those findings. Further, we would like to draw some attention to the little cluster at $(x, y) \sim (-250, -250) \, \mathrm{pc}$ ($l, b \sim 218.5^{\circ}, -2^{\circ}$) of Fig. \ref{fig:fig2}. 
A preliminary inspection of the proper motion and the colour-magnitude diagram (see Appendix C) indicates that this is probably an open cluster, previously unidentified (to the best of our knowledge) due to its proximity to the galactic plane. 
The presence of a new open cluster next to, and possibly related to, the Orion star-forming region adds a new piece to the puzzle of the SFH of Orion.

Some density enhancements are visible only or mostly in the PMS map.
This is because those are low- or intermediate-mass star-forming regions, with very few early type stars.
\begin{enumerate}
\item \textit{Taurus} and \textit{Perseus (Per OB2)}: $x -300 < x < -50 $ and $0 < y < 100 $ pc.
Taurus \citep{Kenyon1994, Scelsi2007} lacks massive OB-type stars and has therefore become a prototype to study  low-mass star-formation processes.
\cite{Belikov2002, Belikov2002b} studied an area of $\sim 20^{\circ}$ diameter centred on the Perseus OB association, identifying over 800 members by their common proper motion and distances. Surprisingly, although even harbouring one of the major associations in the solar vicinity \citep{deZeeuw1999, Bally2008b}, Per OB2 is only barely visible in the UMS map of Fig. \ref{fig:fig21}, probably because of the lower number of massive stars it contains with respect to Orion, Vela, and Sco-Cen.
\item \textit{Cepheus,Cassiopeia,} and \textit{Lacerta (Lac OB1)}: $ -200 < x < -50 $ and $250 < y < 500 $ pc.
Cepheus contains several giant star-forming molecular complexes, located at various distances \citep{Kun2008}. According to their distance they can be arranged in different subgroups: at $d < 500\, \mathrm{pc}$ there are the clouds located in the Cepheus flare  \citep[see Fig. 2 in][]{Kun2008}, while the associations Cep OB2, Cep OB3, and Cep OB4 \citep{deZeeuw1999} are located between $600$ and $900 \, \mathrm{pc}$, and therefore beyond the boundaries of our region. The groups in Fig. \ref{fig:fig2} are associated to the Cepheus flare and follow closely the gas structures. 
Lac OB1 is an association in its final stage of star formation \citep{Chen2008}. The groups that we identified in our maps are:
LBN 437 (also known as Gal 96-15) and Gal 110-13. These are the only regions with recent star-formation activity.
Cassiopeia  contains a few nearby star-forming molecular clouds \citep{Kun2008}. In the maps it is possible to identify a group related to $\mathrm{LkH\alpha} \, 198$ and associated with the dark cloud L 1265, plus another small cluster in the same area.
\item \textit{Aquila}: $ x > 100 $ and $50 < y < 200 $ pc. \\
A few density enhancements are visible towards the Aquila Rift. In general they follow the dust structures, with some small clumps. The density enhancements are not related to the open clusters identified in the UMS map, as the estimated ages of those are older than 20 Myr. We therefore 
conclude that stars in that region of the PMS map are mainly MS contaminants that survive the selection process or are older PMS sources. 
\end{enumerate}

\noindent
A peculiar region is that of \textit{Lyra} and \textit{Cygnus}: $ 0 < x < 200$ and $ 250 < y < 500$.

\noindent
Lyra is predominantly visible in Fig. \ref{fig:fig2}, while Cygnus 
is visible in both Figs. \ref{fig:fig2} and \ref{fig:fig21}, although the density enhancements have a slight offset.
The reason for these differences might be the way in which we select the samples: indeed, we select density enhancements in tangential velocities 
and we then study their density in space; some groups might therefore get lost in the process, especially if they do not stand out significantly with respect to the background. This is further discussed in Section \ref{Section4}.
We note here that Cyg OB4 and Cyg OB7 \cite{deZeeuw1999} are beyond the region studied in this work ($d > 500 \, \mathrm{pc} $). The density enhancements we find lie towards the `Northern Coalsack', towards the Cygnus constellation, and towards the $\delta$ Lyra cluster.
As for Sco OB2, Vel OB2, and Ori OB1, the UMS star distribution is broader than the PMS distribution, and seems to connect different groups.
`We note that, towards the same line of sight, two open clusters are present: Roslund 6 \citep{Roslund1960} and Stock 1 \citep{Osborn2002}. However, they are both too evolved (their age is around 300 Myr) to appear in the PMS maps.

\noindent
By comparing the map contour levels at lower densities, we  further notice that the overall star distribution presents some differences. In particular, the PMS distribution shows a clear gap in the region surrounding the Sun.This is not unexpected, as in the innermost $50-100 \, \mathrm{pc}$ 
groups younger than $20 \, \mathrm{Myr}$ are not present. In the same area
the UMS distribution looks instead smoother, even though the area surrounding the Sun does not contain dense clumps in the distribution (which is consistent with the PMS distribution). This is further discussed in Section \ref{Section4}. 
The overall source distribution in the $X, Z$ plane appears inclined with respect to the galactic plane, however the tilt is dominated by Sco OB2 and Ori OB1. Again, this is further discussed in Section \ref{Section4}.

\noindent
Finally, we note that the maps might look different because different values of $w$ and $k$ were used, however the main features that we described above remain visible for different $k$ and $w$ parameters.

\subsection{Ages of the pre-main sequence sample}
We now study the ages of the PMS sample selected in Section 3.3. During the PMS, younger stars are also brighter. For this reason it is quite straightforward to infer age gradients by studying colour-magnitude diagrams of PMS sources. 

\begin{figure*}
\begin{minipage}{0.5\hsize}
\includegraphics[width=\hsize]{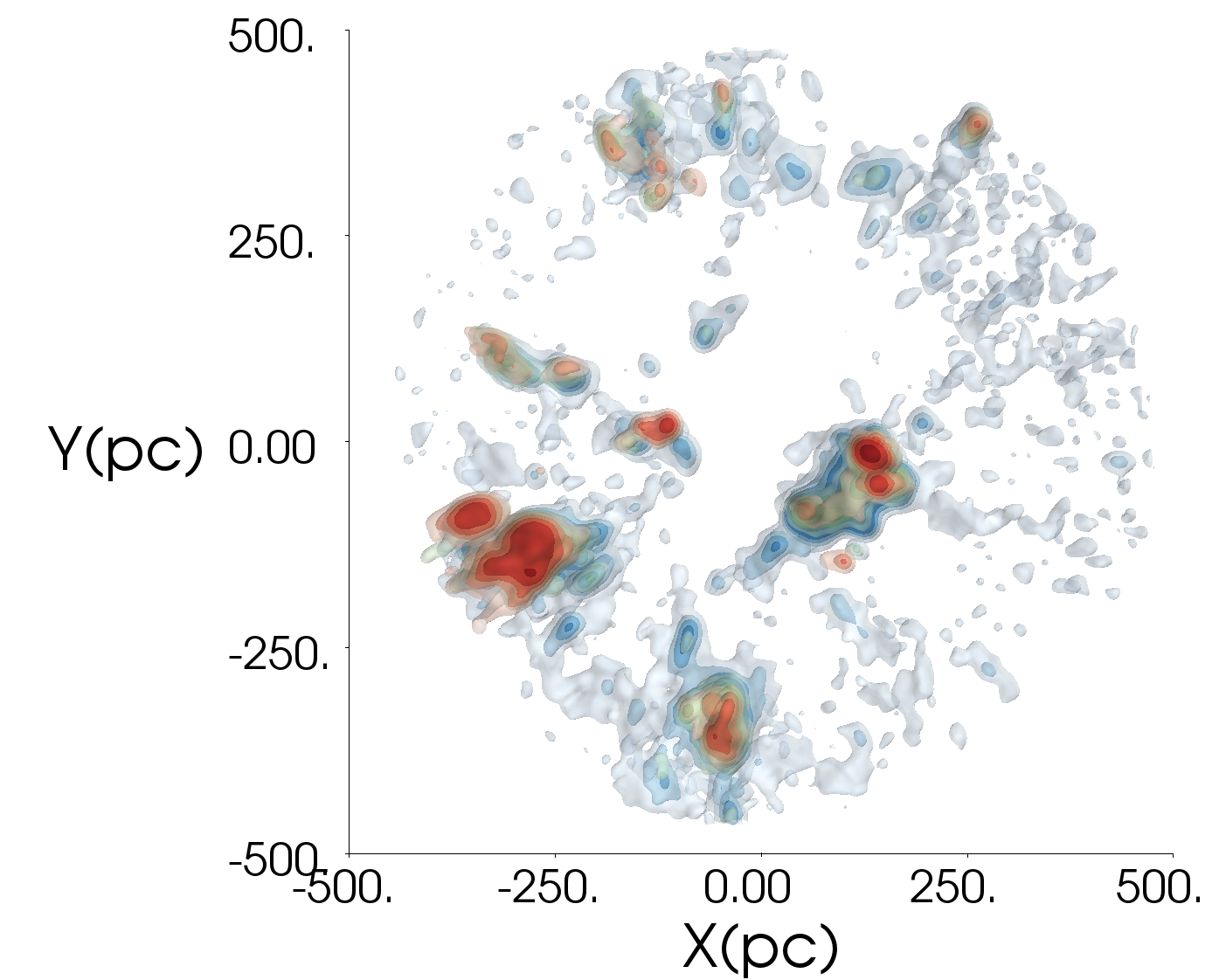}
\end{minipage}
\begin{minipage}{0.5\hsize}
\includegraphics[width=\hsize, height = 0.15\vsize]{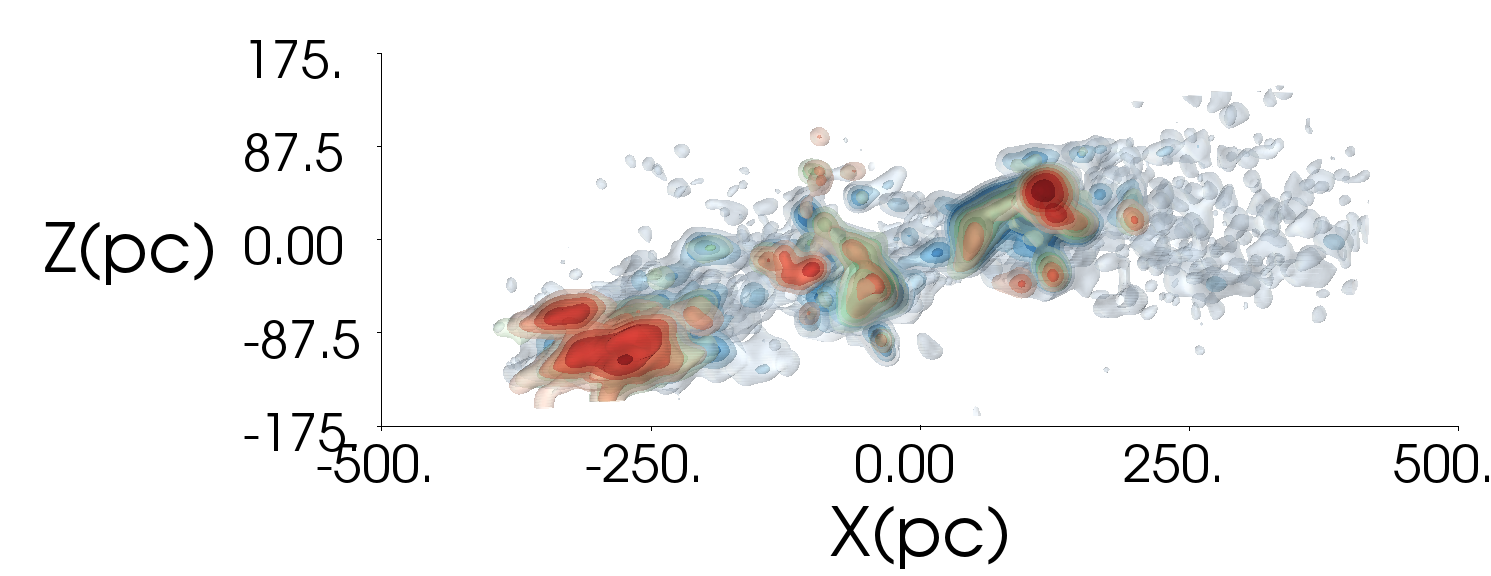}\\
\includegraphics[width=\hsize, height = 0.15\vsize]{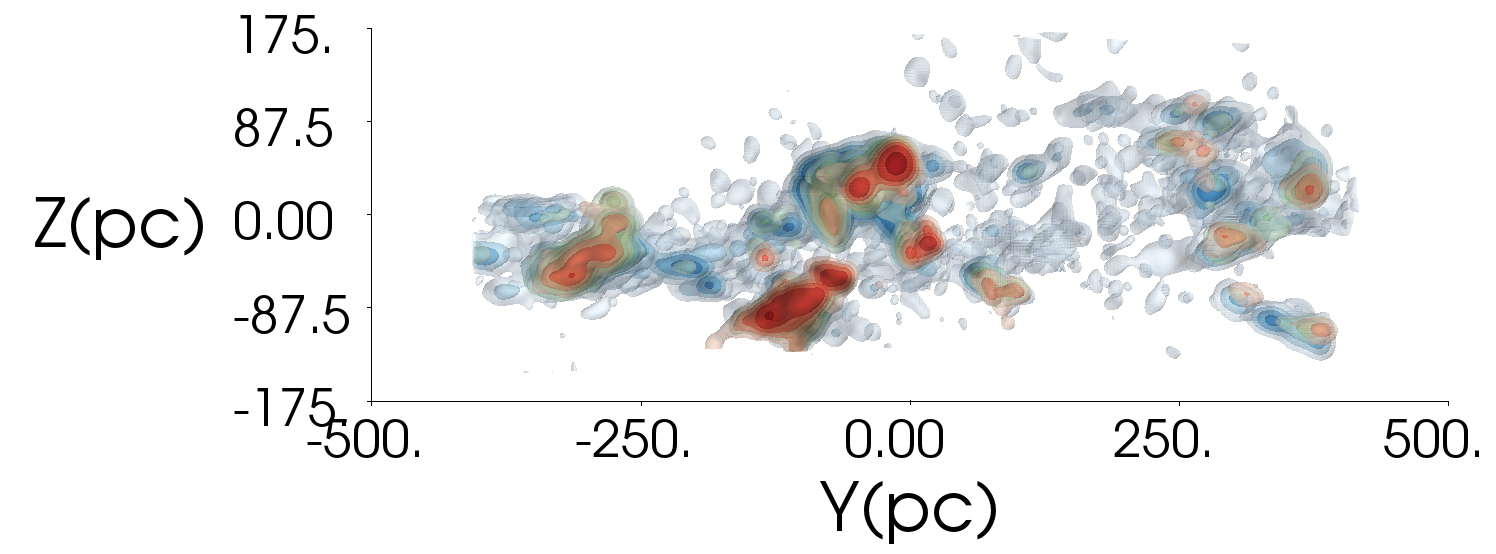}
\end{minipage}
\caption{3D maps of sources younger than 20 Myr and older than 10 Myr (blue), younger than 10 Myr  and older than 5 Myr (green), and younger than 5 Myr (red). The contours are the same as in Fig. \ref{fig:fig2}. In Fig. \ref{fig:appendix2} we show separate maps of the $X-Y$ plane for each of the age intervals.}
\label{fig:fig5}
\end{figure*}

Following the procedure outlined in Section 3.1, we made density maps of the 
PMS stars, dividing them according to their position in the colour-magnitude diagram. We divided the PMS sample into three sub-samples, according to the age ($\tau$) suggested by the PARSEC isochrones: (1) $\tau \le 5 \, \mathrm{Myr}$, (2)
$5 \le \tau \le 10 \, \mathrm{Myr}$,
and (3) $10 \le \tau \le 20 \, \mathrm{Myr}$.
Figure \ref{fig:fig5} shows the density distribution of stars $\le 5 \, \mathrm{Myr}$ (red), $\le 10 \, \mathrm{Myr}$  (green), and $\le 20 \, \mathrm{Myr}$ (blue).
Not unexpectedly the older population is also more dispersed, while younger sources are tightly clustered. The age gradient observed in Sco-Cen by many authors \citep[e.g.][]{Pecaut2016} is evident. In Vela, some young clumps are present, however on average the population is older than in the Orion region. This is not unexpected, as \cite{Jeffries2009} find an age of $\sim 10 \, \mathrm{Myr}$ for the PMS population in Vela. In Perseus, the young cluster IC 348 is visible. The red cluster in $(X, Y) ~  \sim -30, 0 \, \mathrm{pc}$  belongs to the Taurus star-forming region. The groups at large positive $Y$ values are instead more evolved.

\subsection{Caveats}
By performing the source selection that we described in Section 2, we applied different cuts to the data (photometric and astrometric) to clean our sample. In this paper we do not attempt to estimate the purity nor the completeness of the catalogue. The users can make stricter selections based on tangential velocity to obtain a purer sample, at the expense of completeness.

Through extinction mapping we corrected the observed colour-magnitude diagrams and excluded extincted MS sources that contaminated our sample. On one hand, this procedure is necessary to obtain maps that truly trace the distribution of young sources in the solar neighbourhood. On the other hand, the maps might be affected by selection biases introduced by creating the sample, in particular the truncation on relative parallax uncertainty and the application of the extinction correction. \\
\noindent
\textit{Relative parallax uncertainty.} Selecting sources through their relative parallax uncertainty has at least two effects.
\begin{itemize}
\item The ecliptic poles ($ |b| > 45^{\circ}$) are preferred in terms of number of sources due to \textit{Gaia}'s scanning law. This implies that by selecting sources through their relative parallax errors, there  might be a `fake' over-density of sources towards the ecliptic poles (see Appendix \ref{appendix:caveats}). The effect of that would be an over-density in the 3D maps corresponding to those areas or, analogously, an under-density in the other areas. A possible signature of this selection bias might be found in the shape of the low-density contour of the $X-Z$ projection of the  PMS distribution (Fig. \ref{fig:fig2}, right): the density does not look like a uniform slab (compare with the UMS distribution of Fig. \ref{fig:fig21}, right) but presents peculiar `cavities' along $Z$.
This bias - if present - influences the low-density levels and the global source distribution of the maps but not the compact groups that we focus on in this study. 
\item Parallax uncertainties in \textit{Gaia} DR2 increase as a function of increasing $G$ \citep{Brown2018}. Therefore, faint sources at large distances are more easily excluded by the parallax uncertainty selection. This makes our sample incomplete for faint $G$ values. The (in)completeness level is a function of distance (for fixed $G$): for example, a star with $G = 21 \, \mathrm{mag}$ and parallax error $\sigma_\varpi \sim 1 \mathrm{mas}$ \citep[see Fig. 7 in ][]{Brown2018} would be considered part of our sample until $\varpi = 5 \, \mathrm{mas}$ ($d = 200 \, \mathrm{pc}$) and excluded for smaller parallaxes ($d > 200 \, \mathrm{pc}$). While the completeness of the sample needs to be thoroughly analysed when studying the properties of each star-formation region (such as the initial mass function), it should not affect the spatial structures that we observe in the 3D maps.
\end{itemize}
\noindent
\textit{Extinction correction.} While Figs. \ref{fig:fig1} and \ref{fig:fig2112} show essentially a uniform distribution of sources on the galactic plane, without any evident sign of extinction,  Fig. \ref{fig:fig15} (top) clearly shows the outline of nearby molecular clouds. To exclude extincted sources we resolved to eliminate all the PMS sources with $A_G > 0.92 \, \mathrm{mag}$. This cut aims at excluding background, heavily extincted stars, however in practice it removes also 
young stellar objects still embedded in their parental molecular clouds, or actual PMS stars that lie behind a dense cloud (e.g. potential young groups behind the Aquila rift). By comparing the maps of Fig. \ref{fig:fig2} and \ref{fig:appendix3} (where in the latter the condition $A_G  < 0.92 \, \mathrm{mag}$ is not applied), we notice very much the same main density enhancements (see Section 3.2 and Appendix E for more details), and therefore we conclude that the extinction correction that we are applying is satisfactory for our PMS sample, but should not be applied blindly.

\section{Discussion}\label{Section4}
In the previous sections, we analysed the spatial distribution and the age ordering of a young stellar population within $d = 500 \, \mathrm{pc}$ of the Sun. In this section, we discuss our findings in the context of the SFH of the solar neighbourhood.

\begin{figure*}
\includegraphics[width=\hsize]{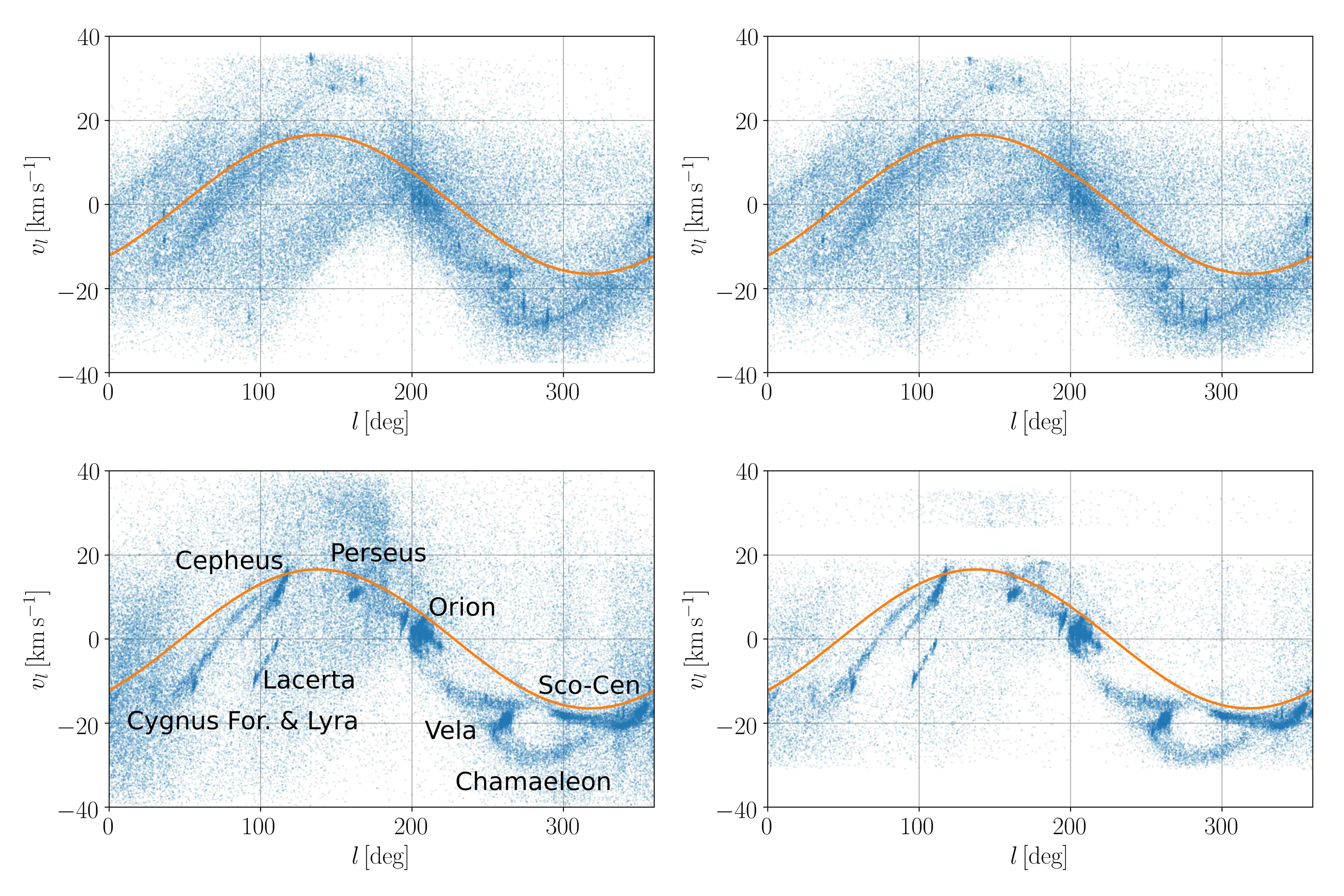}
\caption{Tangential velocity along galactic longitude vs. longitude for the UMS (top) and the PMS (bottom) samples, before (left) and after (right) the tangential velocity selection. The solid orange line shows the projection of the Sun motion. The `gaps' in the scatter plots on the left are due to the tangential velocity selection (see Section 2 in the text).}
\label{fig:fig24}
\end{figure*}

The Gould Belt's definition varies from author to author. It is however striking how we do not find any evidence of a belt-like structure, either for the PMS sample, or for the UMS sample. The tilt observed with respect to the galactic plane is dominated by Ori OB1 and Sco OB2, which are below and above the galactic plane, respectively. This is particularly evident from the $X$ versus $Z$ projections of Figs. \ref{fig:fig2} and \ref{fig:fig5}.
As \cite{Bouy2015} proposed, the existence of a belt of star-forming regions gives a poor description of the spatial distribution of the stars revealed by our analysis, calling for a new interpretation of the distribution of stellar groups in the solar neighbourhood.
Referring to the UMS distribution, we confirm the presence of three large structures, Scorpius-Centaurus, Vela, and Orion, hundreds of parsecs long, which  \cite{Bouy2015} identified and named `blue streams'. The distribution of the PMS stars closely follows the OB distribution and  defines the dense and young regions of the blue streams. 
By using \textit{Gaia} DR2 data, we extend the \cite{Bouy2015} study to include the regions at positive $Y$ values in the maps. Perseus and some clusters in Taurus, as well as Lacerta and Cepheus, are easily visible in our PMS and UMS maps and were not identified by \cite{Bouy2015}, probably because they do not host a large number of early type stars. 
The distributions shown in the maps present some differences: for example, some density enhancements are prominent in only one map. 
As discussed in Section 3.2,  the UMS map shows many open clusters that do not appear in the PMS map because they are older than 20 Myr. In the region corresponding to Taurus we do not observe any density enhancement in the UMS map, as Taurus lacks early-type stars. 
\newline
To further confirm that the main structures that we identify in the PMS map actually  correspond to those in the UMS map, we study the groups in a parameter space that we have not used yet. 
Figure \ref{fig:fig24} shows the tangential velocities along galactic latitude of the UMS (top) and the PMS sample (bottom) older than $20\, \mathrm{Myr}$, before (left) and after (right) the tangential velocity selection of Section 2.2.4. 
The solid orange line shows the projection of the solar motion \citep[$U_{\odot}, V_{\odot}, W_{\odot} = (11.1, 12.24, 7.25) \, \mathrm{km \,  s^{-1}}$,][]{Schonrich2010}. The location of the groups in the $v_l$ versus $l$ plane is primarily due to the projection of the solar motion in different directions. 
The deviations from the solar motion are due to the peculiar motions of the star-forming regions. Clumps and elongated structures are visible, corresponding to the groups mentioned in Section \ref{Section3}.
The features in the PMS panels correspond to those in the UMS panels, although in the latter they are less well defined. Indeed, PMS groups have a smaller velocity dispersion than UMS sources. This agrees with the fact that PMS groups are clustered in denser structures in the 3D maps. Further, by definition, the UMS sample also contains more evolved sources, which are expected to have a larger velocity dispersion. 
The reason for the discrepancies in the maps might therefore be due to the density contrast of different groups. Indeed the stellar population of some groups is more abundant (such as in Sco OB2 or Ori OB1), and/or more compact (in the case of the open clusters observed in the UMS distribution):
the density will peak in these regions, making them stand out more than others. 
Figure \ref{fig:fig24} also shows that the tangential velocity selection is useful to exclude a large number of contaminants, but that  still retains a good number of spurious sources. We note that the gap visible especially in the bottom-right panel of the Fig. \ref{fig:fig24} is due to the tangential velocity selection.
One of the goals of this work is to provide catalogues of PMS and UMS sources  that can be used for future works on the global properties of  solar neighbourhood or on specific star-forming regions. We decided not to impose stricter criteria on our selection to avoid the exclusion interesting sources as much as possible. On the other hand, this means that future users should be careful when using the data, and should combine spatial, kinematic, and photometric data to accurately
select the stellar population of one region.

The most apparent difference in the 3D maps involves the global source distribution. As already noted in Section \ref{Section3}, PMS stars show a gap in their distribution in the inner $\sim 50 \, \mathrm{pc}$. This is not unexpected as the vicinity of the Sun ($d < 50 \, \mathrm{pc}$) is essentially free of stars younger than 20 Myr, except for a few small groups that are difficult to pick up on our maps (e.g. the $\beta$ Pictoris moving group). On the contrary, the distribution of UMS sources appears uniform, with a small under-density next to the Sun that loosely traces the gap observed for the PMS distribution. The fact that the density of early-type stars decreases in the solar vicinity is consistent with the PMS distribution. The distribution is however more uniform for two reasons: the first is related to the smoothing parameters that we used to create the map. Since the number of early-type sources is smaller than that of PMS stars, we had to use a larger value of $\sigma$ to smooth the density distribution (see Section 3.1). The second is related to the age of early-type stars. As we already mentioned above, the UMS consists also of stars whose age is larger than 20 Myr because of the way we selected the sample. For this reason the distribution of the UMS sample is intrinsically more spread out than that of the PMS sample.

\begin{figure}
\includegraphics[width=\hsize]{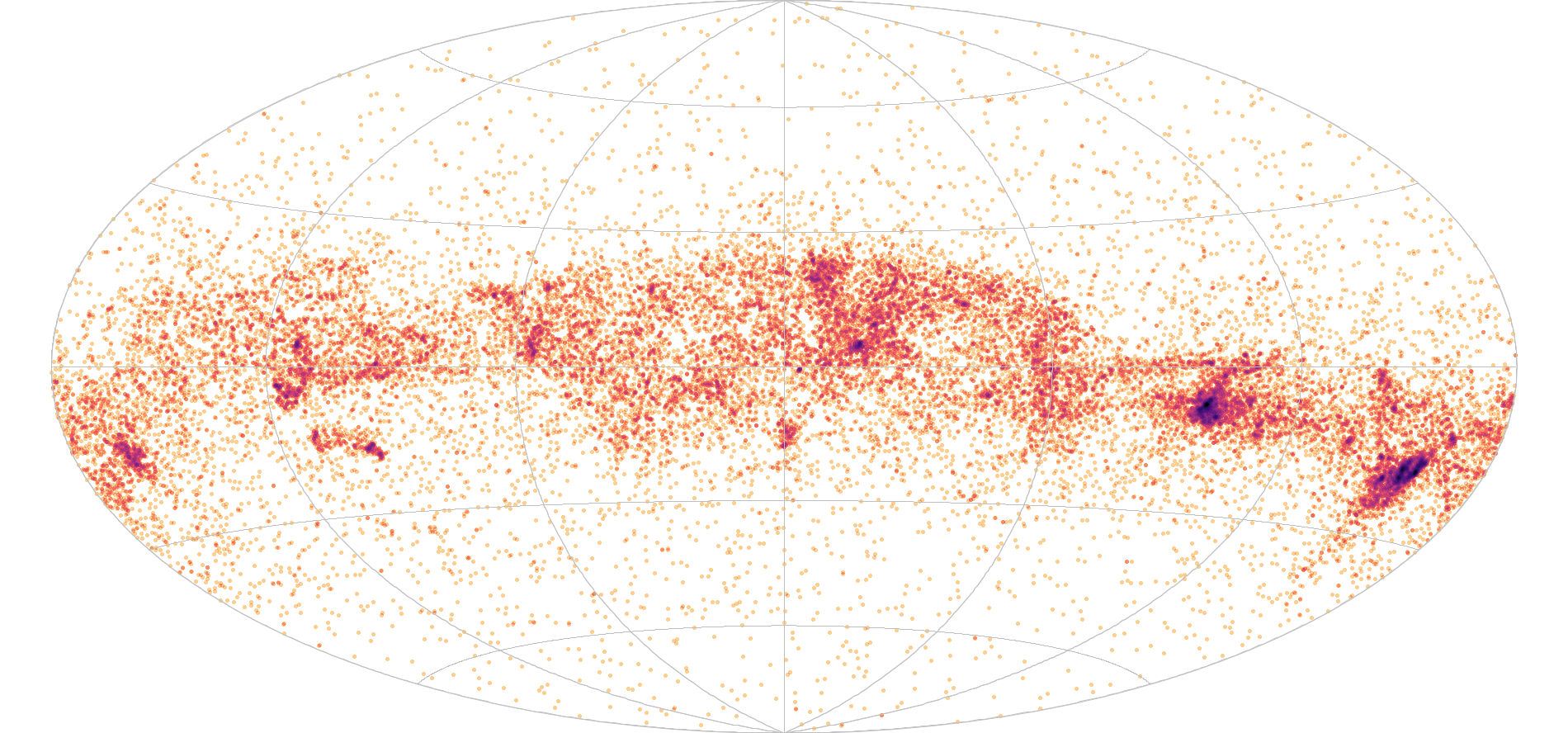}
\includegraphics[width=\hsize]{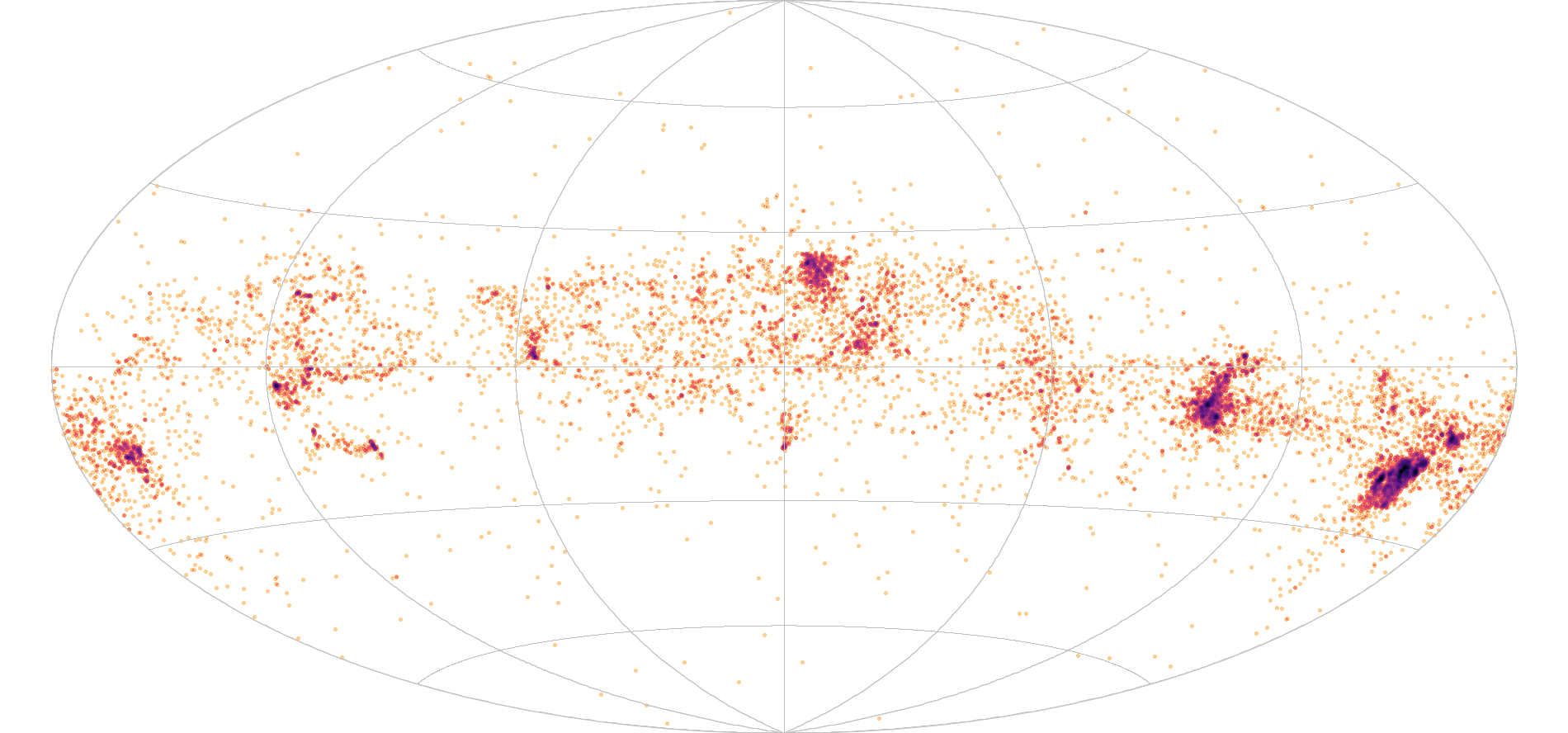}
\includegraphics[width=\hsize]{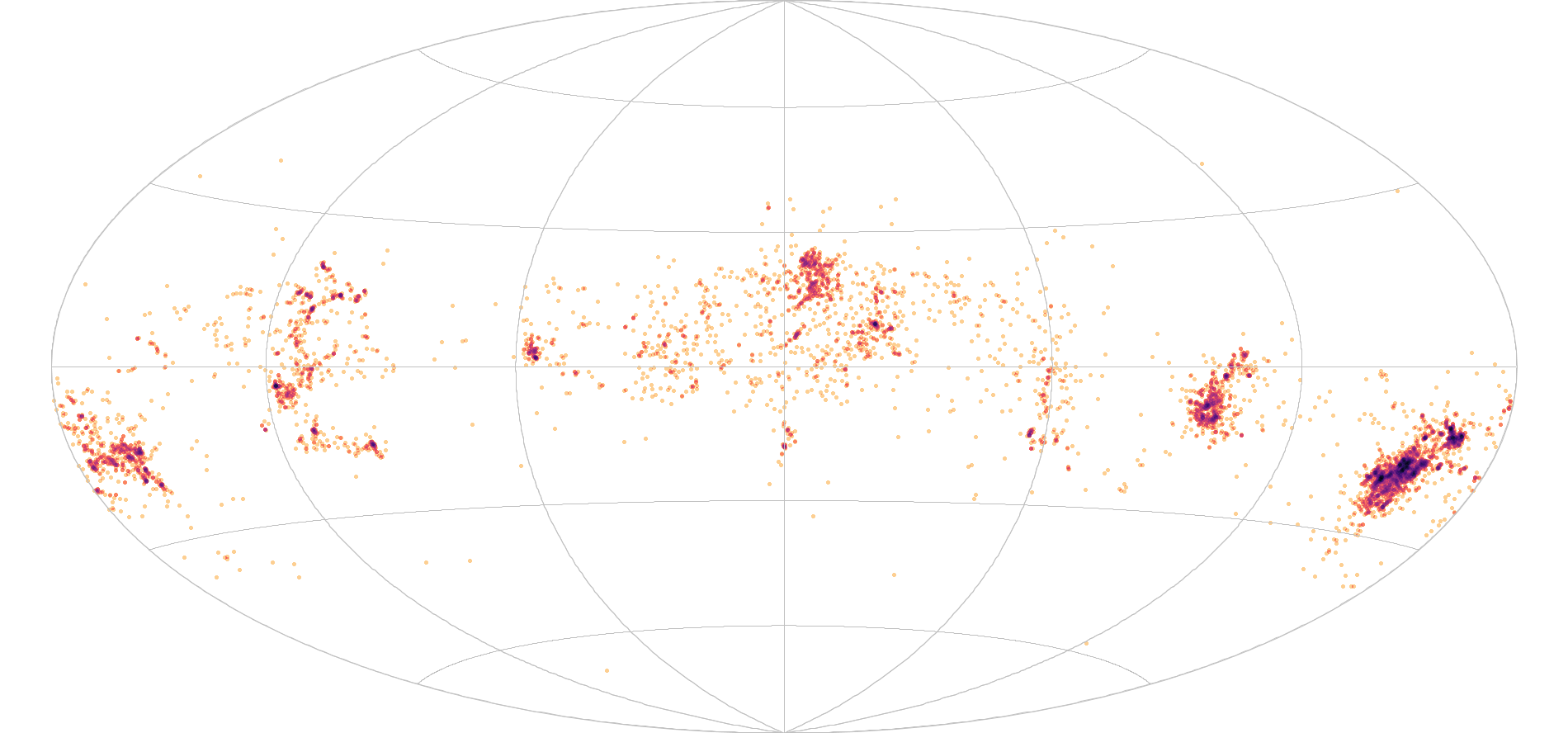}
\caption{Sky projection of sources with different ages. \textit{Top}: sources with $10 < t < 20 \, \mathrm{Myr}$; \textit{centre}: sources with $5 < t < 10 \, \mathrm{Myr}$; \textit{bottom}: sources with $t < 5 \mathrm{Myr}$.}
\label{fig:fig25}
\end{figure}

The age map of Fig. \ref{fig:fig3} suggests that multiple star-formation episodes can occur within the same region and put limits on the duration of a single star-formation episode. 
We notice that a global trend between the different star-forming groups is not present, and that, within each group, older and younger stars are spatially mixed. This is also visible in  Fig. \ref{fig:fig25}, which shows the same sources as in Fig. \ref{fig:fig5}, projected in the sky (older to younger from top to bottom). Younger stars are clustered in denser clumps,
usually surrounded by the older, more diffuse population. 
We note that in our age maps we do not take binarity into account. As discussed in \cite{Zari2017}, unresolved binaries stand out as a separate sequence, which, being brighter by $\sim 0.75 \, \mathrm{mag}$ with respect to the MS, might look like a younger population. This is a major cause of age spreads, and could affect absolute age estimates. However, binarity should affect our data in the same way in all directions and distances, making relative age estimates quite robust.  
In fact, significant age spreads have been observed in young clusters. \cite{DaRio2012} observed an age spread as large as 10 Myr in the Orion Nebula Cluster (ONC). More recently, \cite{Beccari2017} reported three separated PMSs towards the ONC, indicative of three different episodes of star formation, each separated by about 1 Myr.
\cite{Kroupa2018} explained such observations by outlining a scenario where subsequent bursts of star formation are regulated by stellar feedback and dynamical ejections of high-mass stars. 
According to this scenario, after the first episode of star formation, the newly formed stars ionise and suppress star formation in the embedded cluster. However, high-mass stars are soon ejected from the cluster, allowing gas inflow to resume. This sequence of events can be repeated until the maximum lifetime of a molecular cloud (around 10 Myr) is reached. 
Albeit with some stretch of the imagination (the groups we observe in the maps are more extended than the ONC, and the over-densities could encompass more than one cluster), this scenario might also explain our observations: indeed younger groups generally occupy the central regions of the density enhancements and are surrounded by a more diffuse population.\\
The age map also shows age gradients. In Sco OB2, the youngest groups correspond to Upper Scorpius (US), while Upper Centaurus Lupus (UCL) and Lower Centaurus Crux (LCC) \citep[see also][]{Pecaut2016} appear older. In Fig. \ref{fig:fig25} we observe a density enhancement at coordinates $l, b \sim 343^{\circ}, +5^{\circ}$: this cluster has been reported by \cite{Roser2018}, \cite{Villa2018}, and \cite{Damiani2018} and is  traditionally not within the boundaries of Sco OB2. We confirm that given its distance and age, the cluster  is likely related to the association. 
\cite{Krause2018} combined gas observations and hydrodynamical simulations
to study the formation of the Scorpius-Centaurus super bubble, and suggest a refined scenario for the evolution of the OB association. Dense gas is originally distributed in an elongated cloud, which occupies the current area of the association. The star-formation events in UCL and LCC lead to to super-bubbles that expand, surrounding and compressing the parental molecular cloud, triggering star formation in US. This scenario predicts the formation of kinematically coherent sub-groups within the associations that move in different directions, which is similar to the observed kinematics in Sco-Cen \citep{Wright2018}. \cite{Krause2018} also predict that young groups could occur also in regions of older stars, and that several young groups with similar ages might form over large scales. This is consistent with what we observe, not only in Sco-Cen, but also in the other groups. 
In the Orion region, old stars appear to cluster on the sides and in front of the young population (see Fig. \ref{fig:fig24}). The candidate open cluster at $l, b \sim 220^{\circ}, -2^{\circ}$, $X, Y \sim (-250, -250) \, \mathrm{pc}$, has an age  $> 10 \, \mathrm{Myr}$ and might be related to the Orion dust ring discovered by \cite{Schlafly2015}. 
\cite{Cantat2018}  found that young stars in Vel OB2 trace the gas and dust features of the IRAS Vela Shell and proposed that intense supernova activity coming from the Trumpler 10 and NGC 2451B released enough energy to create a cavity and power the expansion of the IRAS Vela Shell, which subsequently produced a shock in the interstellar medium, which then triggered a second burst of star formation. This agrees roughly with what is shown in Fig. \ref{fig:fig25}: young stars in the central panel appear slightly more concentrated in the area corresponding to the shell than older stars in the top panel. This should however be further investigated, as Fig. \ref{fig:fig24} shows an overlap of the sources in the three different age intervals. 
The star-forming regions at positive $Y$ values appear, in general, to be more evolved, and their stellar content is less numerous than that of the groups discussed above. However, as they are located towards well-known and rich star-forming regions, such as the Cepheus and Cygnus OB associations, they might be the extremities of those groups that lie closer to the Sun. This should be further investigated by extending the map out to further distances, but this is beyond the scope of this paper. 

\begin{figure}
\includegraphics[width=\hsize]{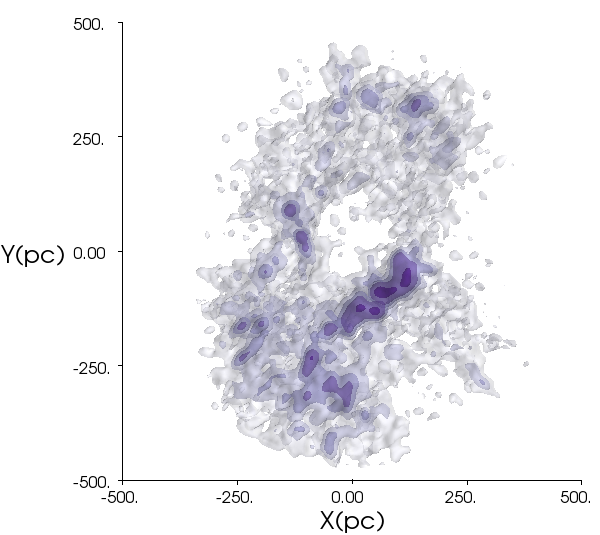}
\caption{3D map of sources older than 20 Myr. The contours represent the 0.2, 0.4, 0.6, 0.8, and 1 density levels.}
\label{fig:fig26}
\end{figure}

Finally, we consider the PMS sources that, according to the isochrones in Fig. \ref{fig:fig12}, are older than 20 Myr, and we select them using the same method outlined in Sections 2.2.2 and 2.2.3.
The spatial distribution of the sources is shown in Fig. \ref{fig:fig26}. The density map presents many interesting features. First, we note that the Orion young population has completely disappeared from the map, while the evolved clusters on its sides are still visible. The Vela and Scorpius-Centaurus populations are still traced by the density distribution, although the density levels appear broader than in the maps of Fig. \ref{fig:fig2}. At positive $Y$ values, the sources related to Cassiopeia, Cepheus, and Chamaeleon are barely visible, however those in the Cygnus foreground and related to the Lyra open cluster are present. This suggests that these regions are relatively evolved, and raises some doubts regarding the connection of the Cygnus foreground to the Cygnus associations. 
The global source distribution is very similar to that presented in the UMS map (Fig. \ref{fig:fig21}). The region surrounding the Sun presents a lack of sources, which is however less pronounced than in the PMS map of Fig. \ref{fig:fig2}. This represents additional evidence that there is a real gap for the youngest stars, extending out to $\sim 100 \, \mathrm{pc}$ towards  Scorpius-Centaurus and reaching $\sim 200 \, \mathrm{pc}$ towards Cygnus and, in the opposite direction, towards Vela and Orion. The gap could therefore be a consequence of any star-forming gas having been cleared out $20-30 \, \mathrm{Myr}$ ago due to the events that created the Local Bubble \citep{Alves2018, Lallement2014, Puspitarini2014}.

\section{Conclusion}\label{Section5}
We used \textit{Gaia} DR2 to study the 3D configuration of early-type, UMS and PMS stars in the solar neighbourhood, within $d = 500 \, \mathrm{pc}$ of the Sun.

\begin{itemize}
\item We selected the data according to a combination of astrometric and photometric criteria. A side product of the data-selection procedure is a 3D $G$-band extinction map which we use to correct our data for extinction and reddening. The final UMS and PMS samples are available online.
\item By using a Gaussian filter smoothing technique, we create 3D density maps for both the UMS and the PMS samples. 
\item The PMS map (Fig. \ref{fig:fig2}) of the sources younger than 20 Myr shows a gap in the innermost $50-100 \, \mathrm{pc}$. 
This is due to the absence of young (with age $< 20$ Myr) groups in the vicinity of the Sun. The same gap also appears in the UMS distribution (Fig. \ref{fig:fig21}), although not as clearly. Due to the way it is constructed, the UMS sample indeed also contains sources older than 20 Myr. This has two effects: 
\begin{enumerate}
\item the low-density distribution appears smoother;
\item more evolved open clusters are visible.
\end{enumerate} 
\item Three structures are discernable in both the maps of Figs. \ref{fig:fig2} and \ref{fig:fig21}: Scorpius-Centaurus, Vela, and Orion. The 
PMS distribution in these regions follows the distribution of the UMS sources, and defines its dense, inner regions. 
\item Taurus, Perseus, Lacerta, Cassiopeia, and Cepheus emerge clearly in the PMS map.
Taurus does not host any young, massive source, therefore it is not visible in the UMS map. Perseus, Lacerta, Cassiopeia, and Cepheus are instead visible as low-level density enhancements. 
\item A peculiar density enhancement is that in the foreground of Cyg OB4 and Cyg OB7: the enhancement is present in both maps, albeit with a slight off-set. We exclude that the PMS density enhancement is related to the open clusters Stock 1 and Roslund 6, as their estimated age is much older 20 Myr. The groups in the foreground of the Cygnus (and Cepheus) associations might therefore represent their extremities that are closer to the Sun.
\item We report the discovery of a young cluster at coordinates $l, b \sim 220^{\circ}, -2^{\circ}$. Due to its position, distance, and age, this cluster might be related to the Orion star-forming complex.
\item We divide the PMS sources into three sub-sets, corresponding to different age ranges ($< 5$ Myr, $5 < t < 10 $ Myr, $10 < t < 20$ Myr),
which we compute by using the PARSEC isochrones. We find that sources in the youngest age sub-sets are more concentrated in space, while those in the oldest age sub-sets are globally more diffuse. Age gradients are visible in many regions, particularly in Scorpius-Centaurus, while in others, such as Vela, stars with different ages appear to overlap in space. 
\item We study the spatial density distribution of the PMS sources older than 20 Myr. At low densities, the density distribution appears similar to the UMS density distribution. The young stellar populations in Orion, Perseus, Cassiopeia, Cepheus, and Chamaeleon are not visible in the map, while Vela and Scorpius-Centaurus are traced by broad density enhancements. At positive $Y$ values, the map shows over-density related to Lyra and to the Cygnus foreground: this implies that those groups are quite evolved and puts into question the relation of the Cygnus foreground to the Cygnus associations.
\end{itemize} 

\noindent
In conclusion, we find that the 3D configuration of the star-forming regions in the solar neighbourhood is far from being described by a ring-like structure such as the Gould Belt, but is complex and filamentary. A detailed analysis is required to precisely order all the star-forming regions according to their ages. In future work we will  
combine \textit{Gaia} data and other spectroscopic surveys to analyse the kinematic properties of the young stars in the solar neighbourhood, something only touched upon here.  The study of the kinematics and internal velocity patterns (such as expansion and contraction) of the concentrations of young stars will provide deeper insight into their origin.

\begin{acknowledgements}
We thank the referee for their constructive comments, which improved the quality of this manuscript.
This project was developed in part at the 2018 NYC Gaia Sprint, hosted by the Center for Computational Astrophysics at the Simons Foundation in New York City. \\
This work has made use of data from the European Space Agency (ESA)
mission {\it Gaia} (\url{https://www.cosmos.esa.int/gaia}), processed by
the {\it Gaia} Data Processing and Analysis Consortium (DPAC;
\url{https://www.cosmos.esa.int/web/gaia/dpac/consortium}). Funding
for the DPAC has been provided by national institutions, in particular
the institutions participating in the {\it Gaia} Multilateral Agreement.
This publication has made use of data products from the Two Micron All Sky Survey, which is a joint project of the University of Massachusetts and the Infrared Processing and Analysis Center/California Institute of Technology, funded by the National Aeronautics and Space Administration and the National Science Foundation. 
This research made use of Astropy, a community-developed core Python package for Astronomy (Astropy Collaboration, 2013). This work has made extensive use of IPython \citep{ipython}, Matplotlib \citep{matplotlib}, astroML \citep{astroML}, scikit-learn \citep{scikit-learn}, and TOPCAT \citep[\url{http://www.star.bris.ac.uk/~mbt/topcat/}]{topcat}. This work would have not been possible without the countless hours put in by members of the open-source community all around the world.   
\end{acknowledgements}

\bibliographystyle{aa}
\bibliography{bibliografy}

\begin{appendix}
\onecolumn
\section{ADQL queries}
We report here an example of the queries used to select the sources in our field and to perform simple cross-matches. \\

\noindent
\textbf{UMS sample: \\} 
\texttt{SELECT *} \\
\texttt{FROM gaiadr2.gaia\_source AS g\\
WHERE g.parallax\_over\_error >= 5 \\
AND g.phot\_g\_mean\_mag + 5 * log10(g.parallax) - 10 <= 4.4 \\
AND g.phot\_bp\_mean\_mag - g.phot\_rp\_mean\_mag <= 1.7 \\
AND g.parallax >= 2.}

\vspace{0.1cm}
\noindent
\textbf{PMS sample: \\}
It is impossible to download all the entries of the catalogue for sources with $\varpi > 2 \, \mathrm{mas}$, and therefore it is necessary to use multiple queries (e.g. the one below) and join the tables afterwards. We also recommend creating an account on the \textit{Gaia} archive.

\vspace{0.2cm}
\noindent
\texttt{SELECT source\_id, l, b, parallax, parallax\_error,  pmra, pmdec, radial\_velocity, pmra\_error, pmdec\_error, radial\_velocity\_error,  phot\_g\_mean\_mag, phot\_bp\_mean\_mag, phot\_rp\_mean\_mag \\
FROM gaiadr2.gaia\_source \\
WHERE parallax >= 2.0 AND parallax <= 2.1}

\section{Source selection based on the relative parallax uncertainty} \label{appendix:caveats}
\begin{figure*}
\includegraphics[width=0.33\hsize]{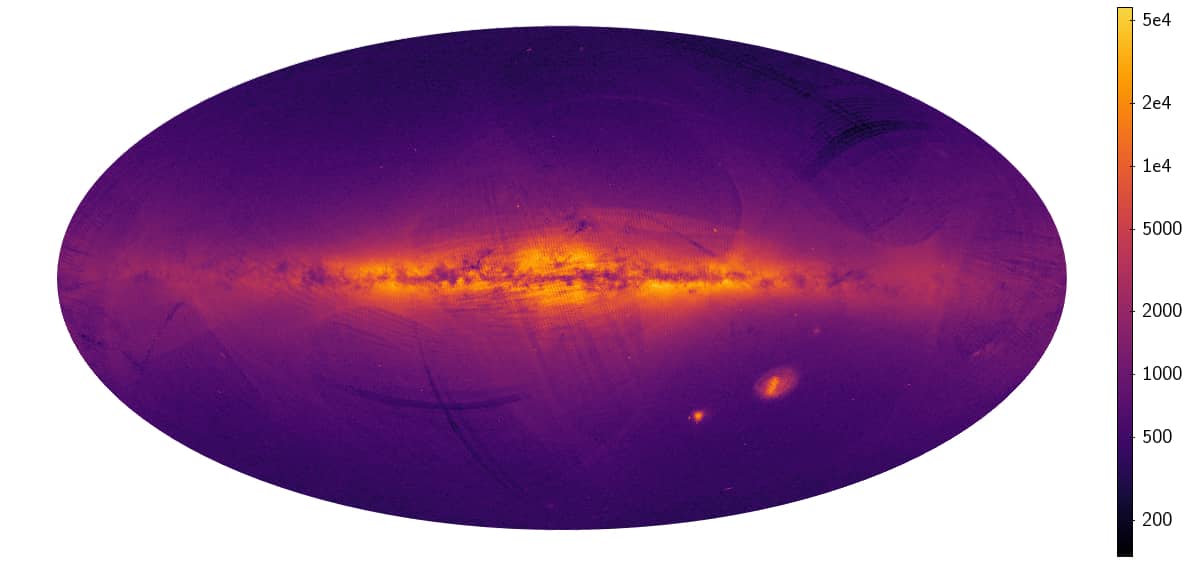}
\includegraphics[width=0.33\hsize]{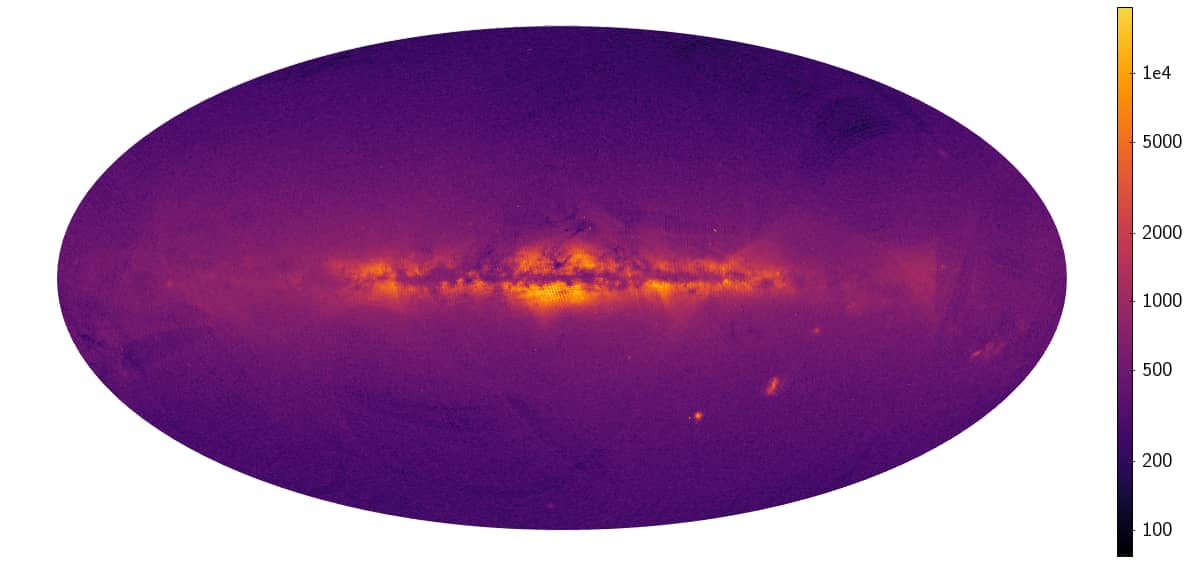}
\includegraphics[width=0.33\hsize]{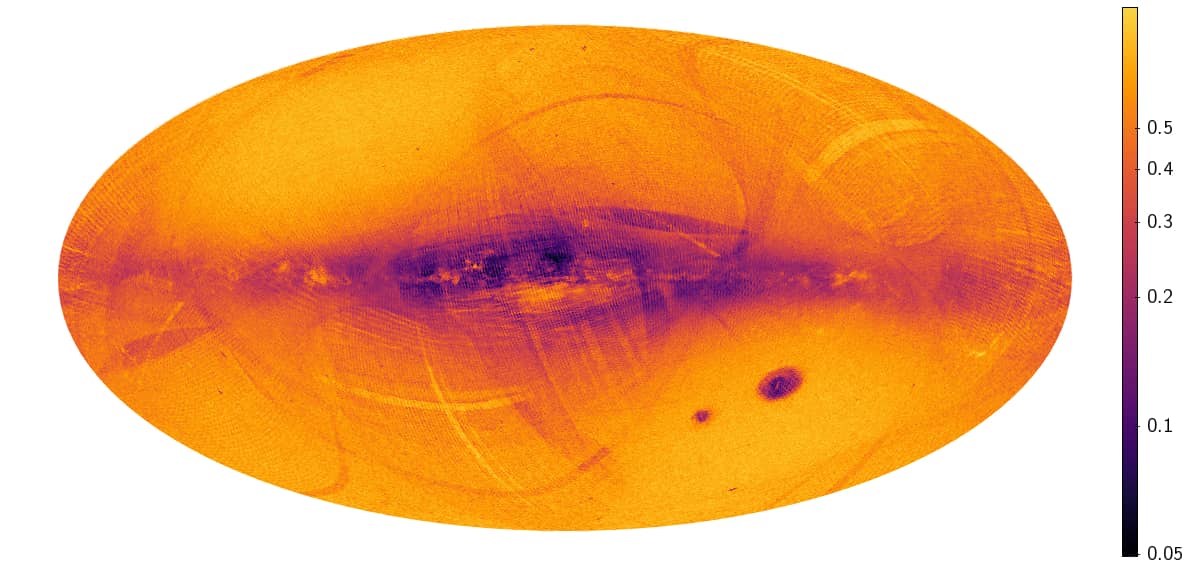}
\caption{\textit{Left}: Distribution in the sky of the sources within $d = 500 \, \mathrm{pc}$. \textit{Centre}: Distribution in the sky of the sources within $d = 500 \, \mathrm{pc}$ and $\sigma_{\varpi}/\varpi > 5$. \textit{Right}: Ratio between the distributions shown in the central and left panels.}
\label{fig:appendix2}
\end{figure*}

In Section 3.4 we mention that by selecting sources basing on their relative parallax errors we might introduce unphysical over-densities in the data due to the fact that \textit{Gaia}'s scanning law favours the ecliptic poles ($|b| > 45^{\circ}$). This effect is easily visible when studying the distribution in the sky of {all} the sources within $d = 500 \, \mathrm{pc}$  before and after applying the condition $\sigma_{\varpi}/\varpi > 5$. Figure \ref{fig:appendix2} (right) shows the ratio between the histograms of the distribution in the sky of the sources before and after the relative parallax uncertainty selection is applied. The ecliptic poles are the regions where the values of the map are close to unity, and without any artefacts due to the scanning law\footnote{Other artefacts are present, such as spuriously high parallaxes: these are taken into account in the text by applying the conditions C.1 and C.2 from \cite{Lindegren2018}.}. The region where we observe the lowest values of completeness is towards the galactic plane for small positive $b$ values. 

\section{New cluster at $l, b \sim (218.5^{\circ}, -2^{\circ})$}
\begin{figure*}
\includegraphics[width=0.33\hsize]{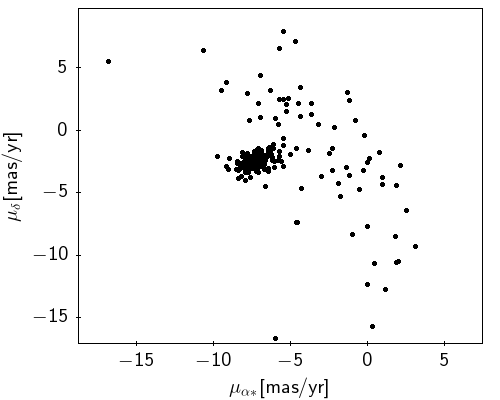}
\includegraphics[width=0.33\hsize]{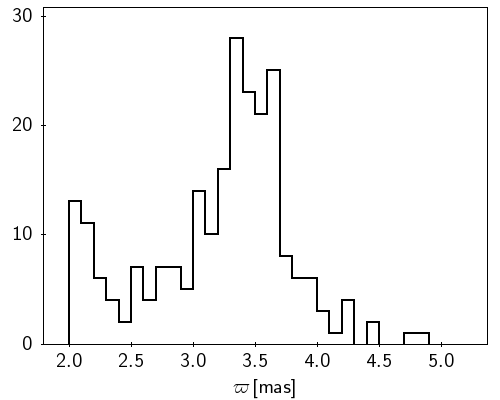}
\includegraphics[width=0.33\hsize]{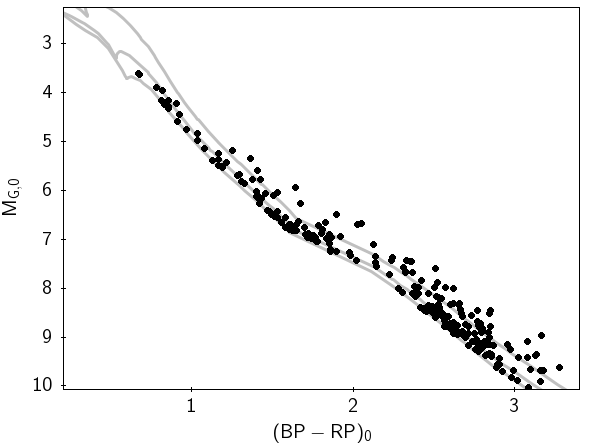}
\caption{\textit{Left}: Proper motion diagram of the sources selected in the region defined in the text. Proper motions cluster at $\mu_{\alpha*}, \mu_{\delta}\sim (-7., -2.5) \, \mathrm{mas \, yr^{-1}}$, with a few scattered outliers.
\textit{Centre}: Parallax histogram of the candidate cluster members. The histogram peaks at $\varpi \sim 3.4 \, \mathrm{mas}$, indicating a distance to the cluster of $\sim 295 \, \mathrm{pc}$. \textit{Right}: Corrected colour-magnitude diagram of the candidate cluster members. The 10, 15, and 20 Myr PARSEC isochrones with solar metallicity and $A_V = 0 \, \mathrm{mag}$ are also plotted as grey solid lines. }
\label{fig:appendix1}
\end{figure*}

As mentioned in the main text of the paper, we report the discovery of a candidate young cluster centred roughly at $l, b = (218.5^{\circ}, -2^{\circ})$. 
Figure \ref{fig:appendix1} shows the proper motion diagram (left), the parallax distribution (centre), and the colour-magnitude diagram (right)
of the sources within $215^{\circ}  \le l \le 222^{\circ}$ and $-5^{\circ} \le b \le 0^{\circ}$.  Except for a few outliers, visible in particular in the proper motion diagram and in the parallax distribution, the cluster prominently stands out as an over-density in the proper motion diagram and as a peak in the parallax distribution.

\section{Age maps}
In this section we separately show the 3D density maps of the sources younger than 20 Myr and older than 10 Myr (blue,  Fig. \ref{fig:appendix2}, right), younger than 10 Myr and older than 5 Myr (green, Fig. \ref{fig:appendix2}, centre), and younger than 5 Myr (red, Fig. \ref{fig:appendix2}, left).
\begin{figure*}
\includegraphics[width=0.33\hsize]{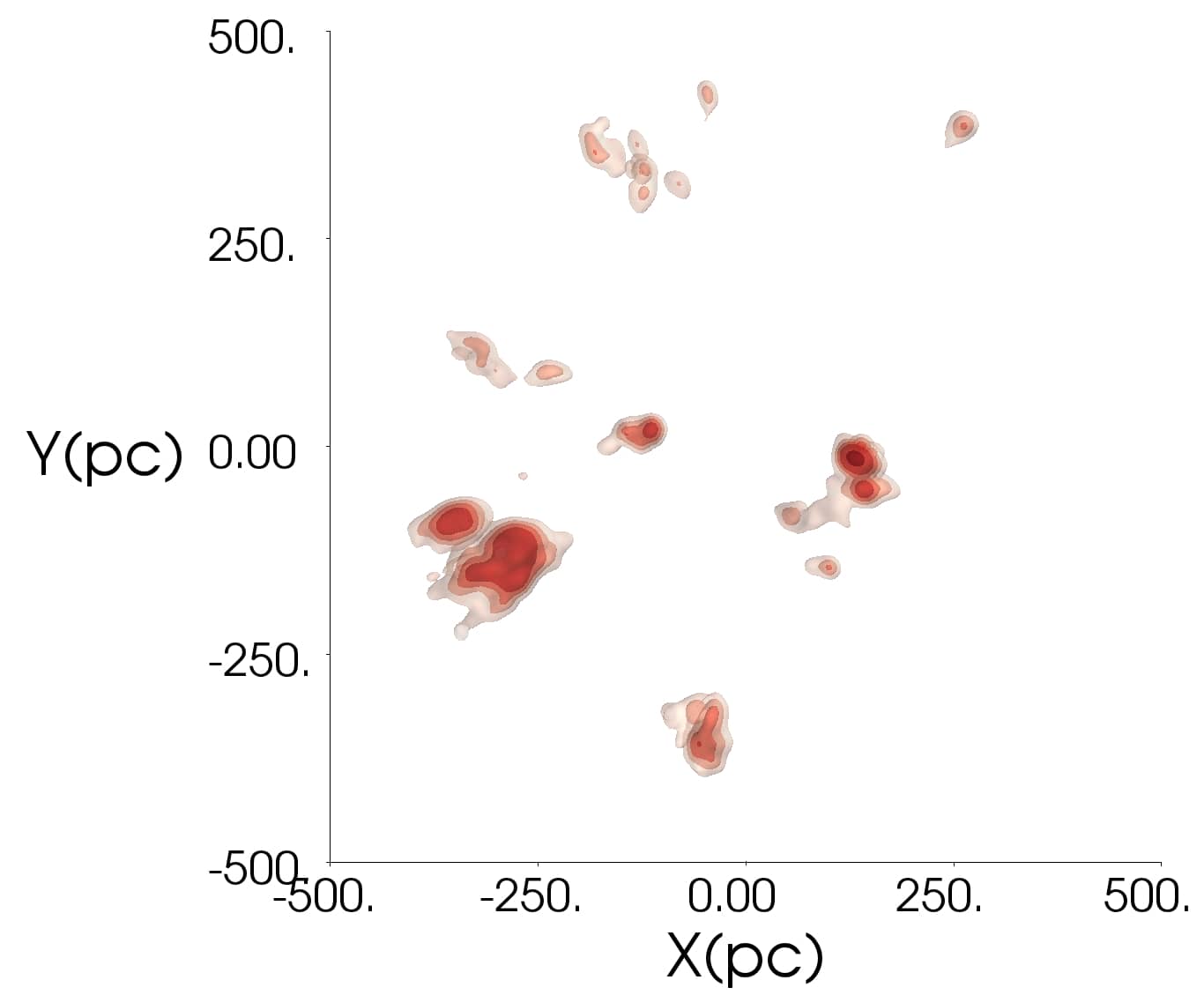}
\includegraphics[width=0.33\hsize]{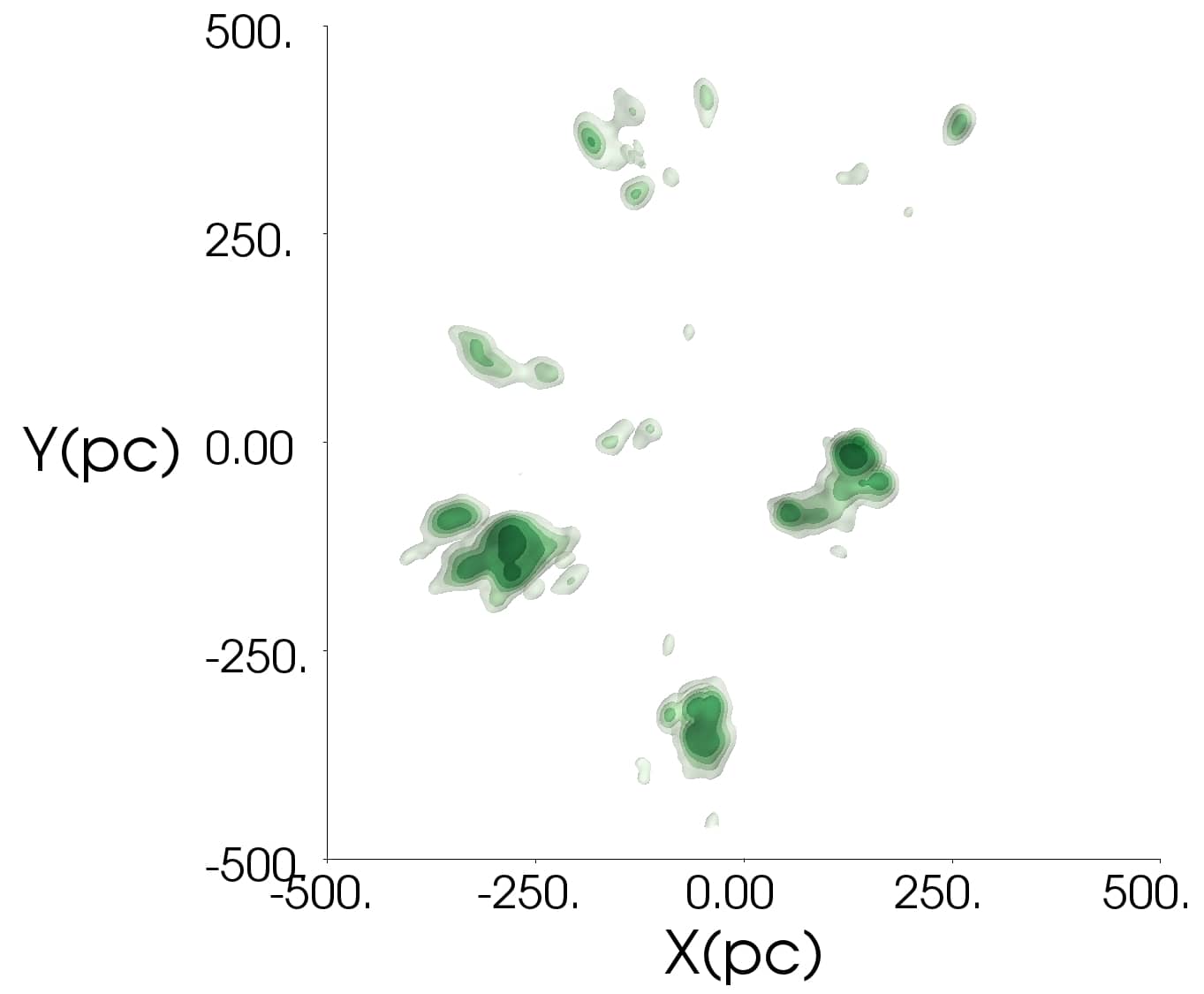}
\includegraphics[width=0.33\hsize]{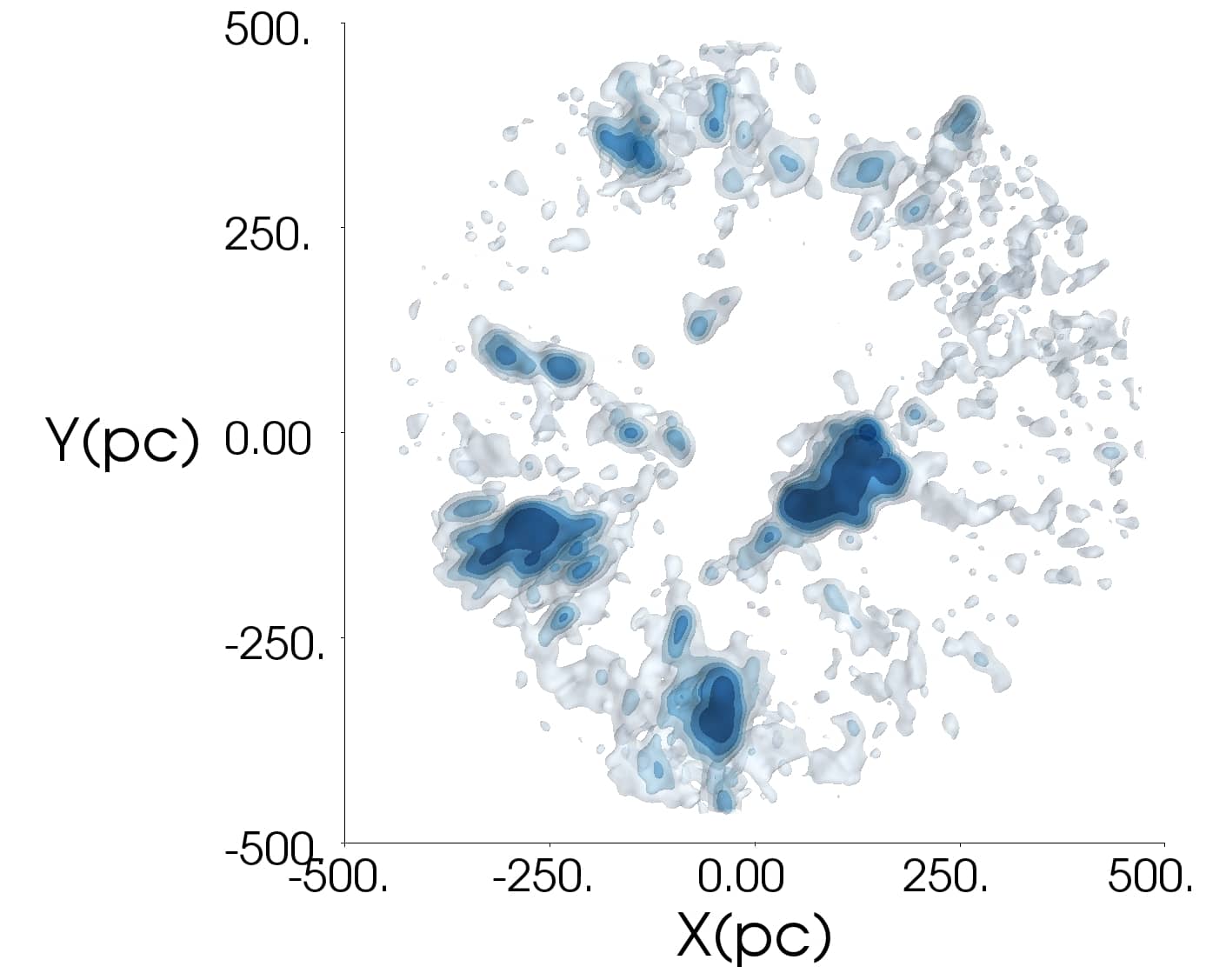}
\caption{3D density map of sources with ages in the ranges $10 < \tau < 20 \, \mathrm{Myr}$ (right), $5 < \tau < 10 \, \mathrm{Myr}$ (centre), and $\tau < 5 \, \mathrm{Myr}$ (left).}
\end{figure*}

\section{Density maps corresponding to the top and central panel of Fig. 6}
\begin{figure}
    \centering
    \includegraphics[width = 0.45\hsize]{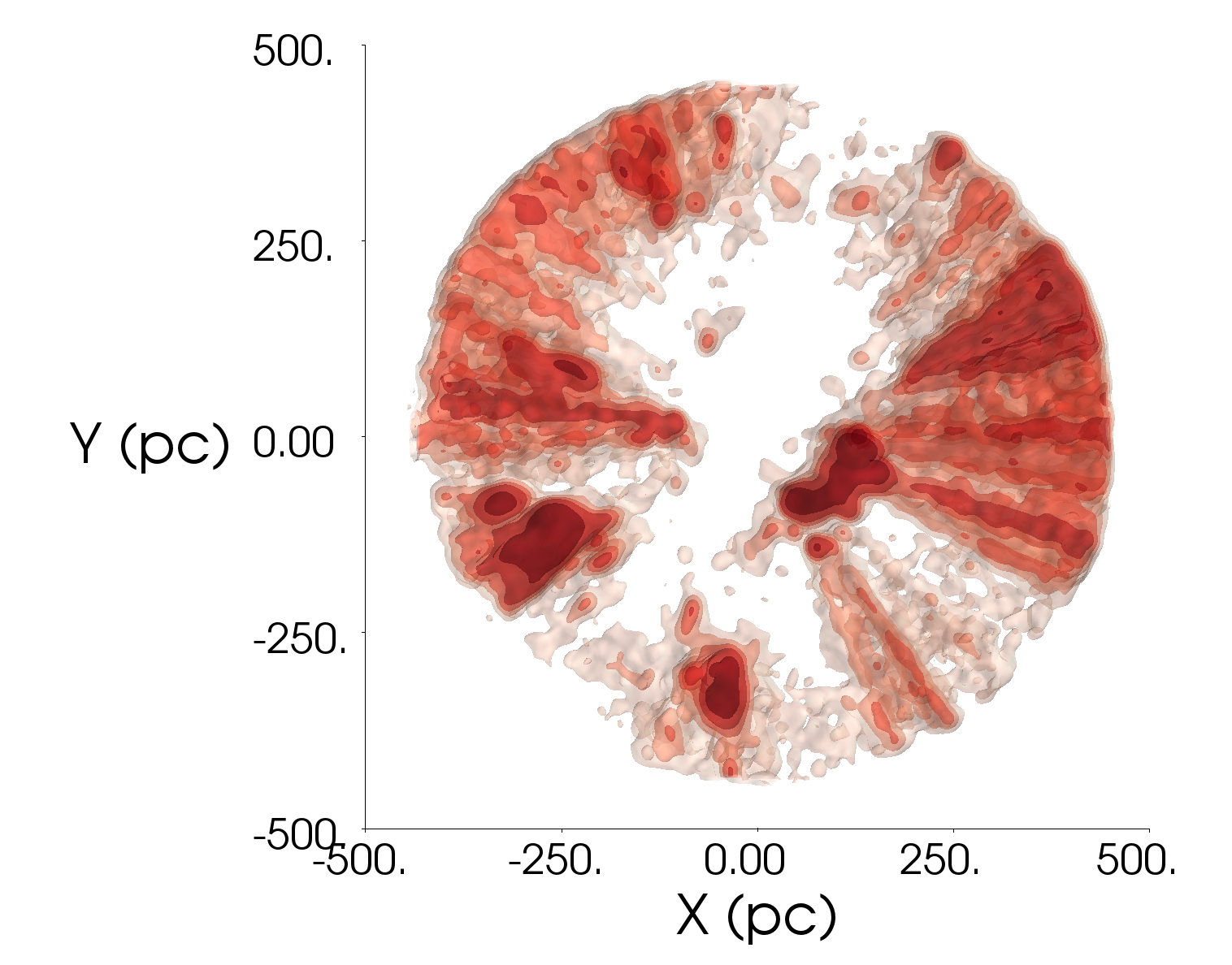}
    \includegraphics[width = 0.45\hsize]{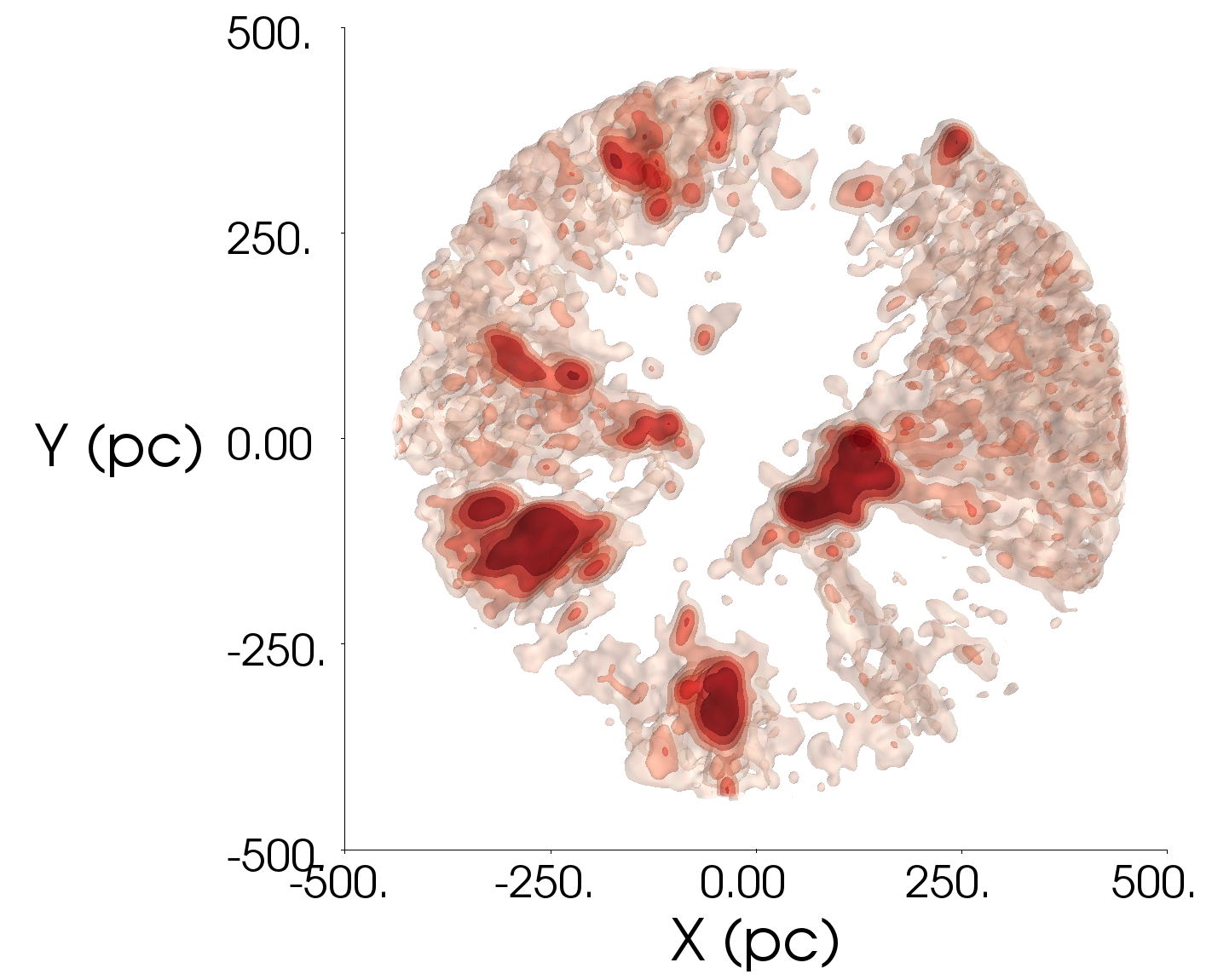}
    \caption{\textit{Left}: 3D density map of the sources in the top panel of Fig. 6. \textit{Right}: 3D density map of the sources in the central panel of Fig. 6.}
    \label{fig:appendix3}
\end{figure}

The conclusion that most of the sources tracing the dust features in the top panel of Fig. 6 correspond to extincted and reddened MS stars, and the subsequent decision to further select PMS candidates according to their extinction and tangential velocity, comes from a preliminary inspection of the 3D density maps. Figure \ref{fig:appendix3} (left) shows the density map corresponding to the top panel of Fig. 6, while Fig. \ref{fig:appendix3} (right) shows the density map corresponding to the central panel of Fig. 6.
Figure \ref{fig:appendix3} (left) does not show any additional clustering with respect to Fig. \ref{fig:appendix3} (right), except for dense `stripes'. These features are located behind molecular clouds \citep[see e.g.][]{Lallement2018}, and they are removed with the condition $A_G < 0.92 \, \mathrm{mag}$, as shown in Fig. \ref{fig:appendix3} (left). Additional contaminants are removed by selecting stars according to their tangential velocity (compare Fig. \ref{fig:appendix3} (right) with Fig. \ref{fig:fig2}).

\section{UMS and PMS catalogues}
Here we briefly describe the contents of the PMS and UMS catalogues.
A detailed description of the column contents and format can be found in the \textit{Gaia} DR2 documentation.
We note that the proper motions are in galactic coordinates, and therefore we provide here the correlation term between proper motion in galactic longitude and proper motion in galactic latitude; we stress however that for a proper use of the \textit{Gaia} DR2 astrometry in galactic coordinates, users should transform the full covariance matrix of the ICRS astrometric parameters.

\begin{itemize}
\item \texttt{source\_id}: unique source identifier (unique within a single release);
\item \texttt{l}: galactic longitude [deg];
\item \texttt{b}: galactic latitude [deg];
\item \texttt{parallax}, parallax [mas];
\item \texttt{parallax\_error}, standard error of parallax [mas];
\item \texttt{pm\_l\_cosb}: proper motion in galactic longitude [mas/yr];
\item \texttt{pm\_l\_error}, standard error of proper motion in galactic longitude [mas/yr];
\item \texttt{pm\_b}:  proper motion in galactic latitude [mas/yr] ;
\item \texttt{pm\_b\_error}: standard error of proper motion in galactic latitude [mas/yr];
\item \texttt{pml\_pmb\_corr}: correlation between proper motion in galactic longitude and proper motion in galactic latitude;
\item \texttt{radial\_velocity}: radial velocity [km/s];
\item \texttt{radial\_velocity\_error}: radial velocity error [km/s];
\item \texttt{phot\_g\_mean\_mag}: G-band mean magnitude  [mag];
\item \texttt{phot\_bp\_mean\_mag}: BP band mean magnitude [mag];
\item \texttt{phot\_rp\_mean\_mag}: RP band mean magnitude [mag];
\item \texttt{phot\_bp\_rp\_excess\_factor}: BP/RP excess factor;
\item \texttt{astrometric\_chi2\_al}: AL chi-square value;
\item \texttt{astrometric\_n\_good\_obs\_al}: number of good observation AL;
\item \texttt{A\_G}: extinction in G-band [mag];
\item \texttt{E\_BPminRP}: colour excess in BP-RP [mag];
\item \texttt{UWE}: Unit Weight Error, as defined in \cite{Lindegren2018}.
\end{itemize}



\end{appendix}
\end{document}